\documentclass[11pt]{article}
\usepackage[a4paper,margin=25mm]{geometry}
\usepackage{amsmath,amssymb,amsthm,mathrsfs,mathtools,array,booktabs,tikz,graphicx}
\usepackage[T1]{fontenc}
\usepackage{lmodern}
\usepackage{microtype}
\usetikzlibrary{arrows.meta,calc,positioning}
\usepackage{cite}
\usepackage{hyperref}
\allowdisplaybreaks[1]\emergencystretch=2em
\newcommand{\dd}{\mathrm{d}}

\makeatletter
\@addtoreset{equation}{section}
\makeatother
\hypersetup{bookmarksdepth=2,colorlinks=true,linkcolor=black,citecolor=black,urlcolor=black}
\hypersetup{pdftitle={Holography for an AdS5-to-Hyperscaling-Violating Flow via Dimensional Reduction from 8d Einstein-Matter Gravity},pdfauthor={Ji-Seong Chae}}

\usepackage{xcolor}

\newcommand{\eps}{\varepsilon}
\newcommand{\AdS}{\mathrm{AdS}}

\begin{document}
\title{Holography for an $\AdS_5$-to-Hyperscaling-Violating Flow\\
via Dimensional Reduction from 8d Einstein--Matter Gravity}
\author{Ji-Seong Chae\thanks{\texttt{jiseongchae17@gmail.com}}\\
Department of Physics, Hanyang University, Seoul 04763, Korea}
\date{}
\maketitle
\begin{abstract}
We construct a classical eight-dimensional Einstein--matter model
whose reduction on a three-torus admits an $\AdS_5$ ultraviolet
vacuum and a hyperscaling-violating infrared solution. Neutral
scalar winding maintains the ultraviolet vacuum, while its
disappearance allows the infrared geometry to become locally
$\AdS_8$. The same reduction fixes the
running gauge coupling, charged scalar mass and interactions.
We derive the vacuum conditions, the radial evolution
equations and the infrared spectrum. Scalar and vector superfluid
orders are then compared using exact ultraviolet zero modes and
analytic infrared eigenvalue identities. The scalar and vector
instabilities in the infrared coincide at a unique scalar mass
above the Breitenlohner--Freedman bound.
Their nonlinear competition
is governed by a quartic free energy that includes electric response
and the change between fixed chemical potential and fixed density.
For the interpolating geometry, we identify the conditions under
which the two orders can appear together and form a locally stable phase.
\end{abstract}
\tableofcontents
\clearpage

\section{Introduction}\label{sec:introduction}
Gauge/gravity duality relates strongly interacting quantum field
theories to classical gravitational dynamics, making observables at
strong coupling accessible through bulk boundary-value problems
\cite{Maldacena:1997re,Gubser:1998bc,Witten:1998qj,Aharony:1999ti}.
At finite density, this correspondence describes thermodynamic and
transport properties of strongly correlated matter in terms of
charged black branes \cite{Hartnoll:2009sz,McGreevy:2009xe}.
Holographic superfluids provide a particularly direct example.
A normalizable charged bulk field represents a spontaneously formed
order parameter, while its nonnormalizable coefficient supplies the
external source \cite{Herzog:2009xv,Horowitz:2010gk,Cai:2015cya}.
Understanding such a phase therefore requires both an equation for
the condensate and a gravitational prescription for comparing the
free energies of the allowed solutions.

The first holographic superfluids realized scalar, or $s$-wave,
condensation near a charged horizon and followed the corresponding
branch into the nonlinear regime
\cite{Gubser:2008px,Hartnoll:2008vx,Hartnoll:2008kx,Franco:2009if}.
An $SU(2)$ gauge field also permits a spatial vector, or $p$-wave,
order parameter
\cite{Gubser:2008wv,Gubser:2008zu,Ammon:2010pg,HerzogPufu}.
Analytic near-critical solutions are available for both scalar and
vector order \cite{Basu:2011tt,Herzog2010}. When both orders are
available, the preferred phase follows from comparing gravitational
saddles with identical boundary conditions \cite{Basu2010}.

The pure $s$, pure $p$ and coexisting $s+p$ phases have been studied
in models with coupled scalar and vector matter, including nonlinear
gravitational backreaction
\cite{Amado:2013splusp,Nie:2013sda,Nie:2014backreaction}.
Related $s+d$ systems exhibit the same distinction between the
appearance and energetic preference of an ordered branch
\cite{Nishida:2014lta,Li:2014wca}.
At a continuous onset, the linearized equation determines when a
normal solution becomes marginal, whereas the quartic free energy
determines whether its continuation forms a mixed phase or gives way
to a pure order \cite{Yin:2013,Natsuume:2025GL}.
The stability of this thermodynamic minimum must also be checked
against density variations and dynamical modes
\cite{Zhao:2023stability,Cao:2024susceptibility,Zhao:2024spinodal}.
These comparisons require consistent scalar boundary conditions,
including the AdS mass bound and alternative quantization where
applicable \cite{BreitenlohnerFreedman:1982,KlebanovWitten:1999}.
Holographic renormalization supplies the corresponding source,
response and finite on-shell action
\cite{Bianchi:2001kw,Skenderis:2002wp}.

The preceding analyses are simplest on asymptotically AdS geometries,
but a strongly coupled theory can change its scaling properties
along the radial direction. Lifshitz and Einstein--Maxwell--dilaton
solutions describe scaling regimes with a running scalar and an
effective dynamical exponent
\cite{Kachru:2008yh,Charmousis:2010zz,Gouteraux:2012yr,Gouteraux:2011ce}.
Hyperscaling violation introduces an additional exponent that
changes the scaling of entropy and spatial observables
\cite{Huijse:2011ef,Dong:2012se,Alishahiha:2012qu}.
For a boundary theory with an AdS ultraviolet fixed point, the
relevant geometry is a radial interpolation from AdS to the
hyperscaling-violating regime. Nonconformal black-brane flows
\cite{GubserNellore:2008} and exact domain walls for particular
Einstein--scalar potentials \cite{Arefeva:2018} make this
interpolation explicit. Because the charged fields propagate over
the entire radial interval, their instabilities depend on the flow
and not only on one of its asymptotic metrics.

This dependence has already appeared in scalar condensation on
AdS-to-hyperscaling-violating backgrounds. The instability analysis
of Cremonini and Li distinguishes an infrared scaling instability
from the full ultraviolet boundary-value problem
\cite{CremoniniLi:2016}. Ren and Xie constructed superconducting
solutions with hyperscaling-violating infrared behavior
\cite{RenXie:2023}, and anisotropic Einstein--Maxwell--dilaton--axion
flows give further examples of distinct infrared scaling endpoints
\cite{Giataganas:2025}.
For vector order, Park, Park and Oh analyzed the dependence on
the dynamical and hyperscaling exponents \cite{Park:2016gzx}.
scalar condensates in the same class of geometries were studied in
Ref.~\cite{Pan:2015hsv}.
More recently, Kim, Choi and Oh developed analytic anisotropic
superfluid solutions near a hyperscaling-violating onset, including
metric backreaction \cite{Kim:2025hsv}.
These results show why the scalar and vector eigenfunctions must be
computed on a common flow if their nonlinear competition is to be
compared at fixed chemical potential or fixed density. Entanglement
entropy provides a complementary geometric probe of these backgrounds
\cite{Ryu:2006bv,Ryu:2006ef,Casini:2011kv,Bhattacharya:2012mi},
including the orientation dependence studied in holographic
superfluids \cite{Cai:2012nm,Park:2022oek,Chen:2024isotropy}.

There is a further issue beyond solving the charged radial equations.
Multicritical behavior depends on how the scalar, vector and neutral
sectors interact \cite{Martini:2018}, yet the potential and gauge
couplings of a five-dimensional Einstein--matter theory are often
specified independently. Dimensional reduction offers a way to
derive their relative normalizations from a single parent action.
Gout\'eraux, Smolic, Smolic, Skenderis and Taylor related
Einstein--Maxwell--dilaton holography to higher-dimensional AdS
gravity and extracted its thermodynamics and holographic dictionary
\cite{GouterauxSmolic:2012,Smolic:2013}.
Reductions including higher-derivative terms, coupled perturbations
and hyperscaling-violating backgrounds further demonstrate the
role of the parent geometry
\cite{Charmousis:2012reduction,Gath:2013,Perlmutter:2012,CremoniniSinkovics:2014}.
The Q-lattice mechanism permits scalar phases that depend on an
internal or spatial coordinate while their stress tensor remains
homogeneous \cite{DonosGauntlett:2014}.

The higher-dimensional perspective suggests a more specific question
for the present flow. Can an $\AdS_5$-to-hyperscaling-violating
geometry arise by reducing an eight-dimensional parent whose
infrared region is locally $\AdS_8$? A reduction on three compact
directions offers a concrete setting for this possibility, provided
that the parent also fixes the interactions of the charged fields.
Related studies of $\AdS_8$ have addressed different aspects of
eight-dimensional gravity. C\'ordova, De Luca and Tomasiello
constructed non-supersymmetric $\AdS_8$ solutions in massive type IIA
supergravity \cite{CordovaDeLucaTomasiello:2019}, while Anastasiou
and collaborators analyzed holographic renormalization of
asymptotically $\AdS_8$ Einstein gravity
\cite{AnastasiouAraya:2026}. These results provide distinct
higher-dimensional settings, but neither supplies the reduction and
charged phase dynamics considered here. The ultraviolet
holographic counterterms in our construction remain those of the
five-dimensional $\AdS_5$ boundary theory.
The relation between higher-dimensional geometry and the phases of
the reduced theory has received renewed attention through
uplift-based classifications and first-order radial descriptions
\cite{GKN:2024,KKN:2025,GKN:2026}.
Related analyses treat entanglement and light scalars across
geometric transitions \cite{JainJarvinen:2025,Faedo:2024}.
For nonconformal D$p$-brane systems, dimensionally continued AdS
descriptions organize correlation functions \cite{Bobev:2025Dp},
while recent supergravity studies further develop scale-covariant
solutions and radial flows \cite{Bobev:2026Scale,Conti:2026D2}.

These developments lead to a specific question. Can one obtain an
AdS-to-hyperscaling-violating interpolation and its charged
interactions from one finite-dimensional Einstein--matter action,
and then use the resulting couplings to compare competing ordered
saddles in both thermodynamic ensembles? The first requirement is
stronger than reproducing a dilaton exponent, since the same
reduction must also determine the neutral potential, charged mass
and quartic coupling. The second requirement is stronger than finding
two coincident linear zero modes, since an $s+p$ solution must be
compared with both pure orders after the electric and gravitational
fields have responded.

We study these questions in a classical eight-dimensional
Einstein--matter theory, retaining the neutral fields that determine
the geometry together with a charged scalar and a Yang--Mills field.
We first derive the geometric flow and the charged interactions from
the same action. We then use exact ultraviolet profiles to compare
the free energies of the scalar and vector orders. The infrared
analysis shows how their onset scales depend on the parent theory.
Finally, we formulate the separate conditions needed for a mixed
phase on the complete flow. The endpoint calculations do not, on
their own, establish that the two complete-flow onset curves cross.

Section~\ref{sec:reduction} derives the higher-dimensional action,
the five-dimensional reduction and the neutral radial equations.
Section~\ref{sec:charged} develops the charged theory and its exact
conformal reference calculation. Section~\ref{sec:phase} analyzes
infrared onset operators, the local mixed-order functional and
its higher-dimensional interpretation. Section~\ref{sec:discussion}
discusses the physical scope. The appendices give reduction
identities, radial responses, neutral optical response and
solvability conditions.

\section{Einstein--matter reduction and the AdS-to-HSV flow}\label{sec:reduction}
The geometric interpretation of the flow requires a parent action
and an explicit relation between its fields and the five-dimensional
variables. We begin with the higher-dimensional theory and its
reduction ansatz, keeping the neutral and charged sectors in the
same action. The reduced action supplies the vacuum and radial
field equations, whose boundary conditions we specify in the same
conventions. This fixes the setting in which the charged condensates
are to be studied.

\subsection{Eight-dimensional Einstein--matter action}
We consider a classical eight-dimensional Einstein--matter theory
with a negative cosmological constant. Upper-case indices
$M,N$ run over the
eight bulk coordinates. The five external coordinates are $x^\mu$,
labeled by Greek indices, and the three periodic internal coordinates
are $y^I$ with $I=1,2,3$. The adjoint index $a=1,2,3$ labels the
$SU(2)$ generators. The ultraviolet $\AdS_5$ radius $L_5$ is the unit
of length. The charged Yang--Mills--doublet density is denoted by
$L_{\rm ch}$, and its strength relative to the neutral gravitational
sector is $\eps$. The parent action is
\begin{align}
S_8&=\frac{1}{2\kappa_8^2}\int\dd^8x\,\sqrt{-\widehat g}
 \left[R_8+\frac{42}{L_8^2}
 -\sum_{I=1}^3\left(|\partial\Psi_I|^2+m_\Psi^2|\Psi_I|^2
                  +\frac{\lambda_\Psi}{2}|\Psi_I|^4\right)
 +\eps\,L_{\rm ch}\right], \label{eq:action}\\
L_{\rm ch}&=-\frac14 G^a_{MN}G^{aMN}
-(D_M\boldsymbol\Phi)^\dagger D^M\boldsymbol\Phi
-\left(m_0^2+\xi R_8+\eta\sum_I|\Psi_I|^2\right)
 \boldsymbol\Phi^\dagger\boldsymbol\Phi
-\lambda_8(\boldsymbol\Phi^\dagger\boldsymbol\Phi)^2 . \label{eq:charged}
\end{align}
Here $\widehat g_{MN}$, $\widehat g$ and $R_8$ denote the parent metric,
its determinant and its Ricci scalar, and $\kappa_8^2=8\pi G_8$.
The vacuum Einstein equation of $R_8+42/L_8^2$ admits $\AdS_8$
of radius $L_8$. Each complex neutral scalar $\Psi_I$ has mass
$m_\Psi$, self-coupling $\lambda_\Psi$ and an independent global
$U(1)$ phase symmetry. The charged fields are an $SU(2)$ connection
$B_M^a$ and a fundamental doublet $\boldsymbol\Phi$ with mass
parameter $m_0^2$, curvature coupling $\xi$, neutral portal $\eta$
and quartic coupling $\lambda_8$. We take $\eps>0$, with the
probe approximation defined by $\eps\to0$. Using Pauli matrices
$\sigma^a$, $\epsilon^{123}=1$ and $\tau^a=\sigma^a/2$, we fix
\begin{equation}
 B_M=B_M^a\tau^a,\qquad D_M\boldsymbol\Phi=(\partial_M+iB_M)\boldsymbol\Phi,
 \qquad G_{MN}=\partial_MB_N-\partial_NB_M+i[B_M,B_N].
 \label{eq:gauge_conventions}
\end{equation}
In components, $G^a_{MN}=\partial_MB_N^a-\partial_NB_M^a
-\epsilon^{abc}B_M^bB_N^c$. Grouping the curvature terms of
\eqref{eq:action} and \eqref{eq:charged} gives
\begin{equation}
 R_8-\eps\xi R_8\boldsymbol\Phi^\dagger\boldsymbol\Phi
 =\mathcal F R_8,\qquad
 \mathcal F=1-\eps\xi\,\boldsymbol\Phi^\dagger\boldsymbol\Phi .
 \label{eq:F_definition}
\end{equation}
The reduction below is performed on the complete action, including
the $\mathcal F R_8$ term. Its metric variation gives
\begin{equation}
 \mathcal F\left(R_{MN}-\frac12\widehat g_{MN}R_8\right)
 +(\widehat g_{MN}\Box_8-\nabla_M\nabla_N)\mathcal F
 =\frac12\left(T_{MN}^{(0)}+\eps T_{MN}^{({\rm ch,min})}\right),
 \label{eq:8dEinstein}
\end{equation}
where $T_{MN}^{(0)}$ is the stress tensor of the cosmological and
neutral-scalar terms and $T_{MN}^{({\rm ch,min})}$ that of
$L_{\rm ch}$ with the curvature coupling omitted. For the neutral
sector the individual terms are
\begin{align}
 T_{MN}^{(0)}={}&\frac{42}{L_8^2}\widehat g_{MN}
 +\sum_{I=1}^3\Big[2\operatorname{Re}
 (\partial_M\Psi_I^*\partial_N\Psi_I)
 -\widehat g_{MN}\big(|\partial\Psi_I|^2
 +m_\Psi^2|\Psi_I|^2+\tfrac{\lambda_\Psi}{2}|\Psi_I|^4\big)\Big].
 \label{eq:8d_neutral_stress}
\end{align}
The corresponding charged stress tensor, excluding the curvature
term already present on the left of \eqref{eq:8dEinstein}, is
\begin{align}
 T_{MN}^{({\rm ch,min})}={}&
 G^a_{MP}G_N^{aP}-\frac14\widehat g_{MN}G^a_{PQ}G^{aPQ}
 +2\operatorname{Re}\!\left[(D_M\boldsymbol\Phi)^\dagger
 D_N\boldsymbol\Phi\right]\nonumber\\
 &-\widehat g_{MN}\left[(D_P\boldsymbol\Phi)^\dagger
 D^P\boldsymbol\Phi+
 (m_0^2+\eta\sum_I|\Psi_I|^2)
 \boldsymbol\Phi^\dagger\boldsymbol\Phi
 +\lambda_8(\boldsymbol\Phi^\dagger\boldsymbol\Phi)^2\right].
 \label{eq:8d_charged_stress}
\end{align}
The two scalar variations yield
\begin{align}
 \Box_8\Psi_I
 -\big(m_\Psi^2+\lambda_\Psi|\Psi_I|^2
       +\eps\eta\,\boldsymbol\Phi^\dagger\boldsymbol\Phi\big)\Psi_I&=0,
 \nonumber\\
 D_MD^M\boldsymbol\Phi-
 \big(m_0^2+\xi R_8+\eta\sum_I|\Psi_I|^2
       +2\lambda_8\boldsymbol\Phi^\dagger\boldsymbol\Phi\big)
 \boldsymbol\Phi&=0 .
 \label{eq:parent_scalar_equations}
\end{align}
We will use the internal components of \eqref{eq:8dEinstein}
to verify the reduction, the neutral equations to find its
vacua and radial flow, and the trace of \eqref{eq:8dEinstein}
to evaluate the nonminimal curvature term in the nonlinear
charged equations. The Yang--Mills equation is obtained by varying
\eqref{eq:charged} with respect to $B_M^a$. Its internal projection
is checked after the toroidal ansatz is inserted.

\subsection{Toroidal reduction to five dimensions}
The diagonal metric ansatz retains the torus breathing mode and the
external metric. It is the specialization of the ordinary $D=d+n$
Einstein-frame ansatz derived in Appendix~\ref{app:general_reduction}.
For the present fixed torus we set
\begin{equation}
 \dd s_8^2=X\,\dd s_5^2+X^{-1}\sum_{I=1}^3\dd y_I^2,\qquad
 X=e^{\phi/\sqrt3},\qquad
 \Psi_I=f(x)e^{iky^I},\qquad B_I^a=0,\quad\partial_I\boldsymbol\Phi=0.
 \label{eq:uplift}
\end{equation}
The external metric $g_{\mu\nu}$ is in five-dimensional Einstein frame.
$\phi$ is the canonically normalized torus volume modulus and $f$
the common winding amplitude. We take $y^I\sim y^I+2\pi/|k|$.
The metric factors are
\begin{equation}
 \widehat g^{\mu\nu}=X^{-1}g^{\mu\nu},\qquad
 \widehat g^{IJ}=X\delta^{IJ},\qquad
 \sqrt{-\widehat g}=X\sqrt{-g},\qquad
 R_8=X^{-1}\left[R_5-\frac12(\partial\phi)^2
                         -\frac1{\sqrt3}\Box_5\phi\right],
 \label{eq:Ricci}
\end{equation}
with $R_5$ and $\Box_5$ formed from $g_{\mu\nu}$. The gauge and
doublet densities follow by inserting these factors,
\begin{equation}
 \begin{aligned}
 \sqrt{-\widehat g}\,G^a_{\mu\nu}G^{a\mu\nu}
 &=\sqrt{-g}\,X^{-1}G^a_{\mu\nu}G^{a\mu\nu},&
 \sqrt{-\widehat g}\,(D_\mu\boldsymbol\Phi)^\dagger D^\mu\boldsymbol\Phi
 &=\sqrt{-g}\,(D_\mu\boldsymbol\Phi)^\dagger D^\mu\boldsymbol\Phi,\\
 \sqrt{-\widehat g}\,(\boldsymbol\Phi^\dagger\boldsymbol\Phi)^j
 &=\sqrt{-g}\,X\,(\boldsymbol\Phi^\dagger\boldsymbol\Phi)^j
 \quad(j=1,2),&
 \sqrt{-\widehat g}\,\sum_I|\partial_I\Psi_I|^2
 &=\sqrt{-g}\,3k^2X^2f^2 .
 \end{aligned}
 \label{eq:reduced_densities}
\end{equation}
The corresponding neutral densities are
\begin{equation}
 \begin{aligned}
 \sqrt{-\widehat g}\sum_I|\partial_\mu\Psi_I|^2
 &=3\sqrt{-g}\,(\partial f)^2,\\
 \sqrt{-\widehat g}\sum_I
 \left(m_\Psi^2|\Psi_I|^2+\frac{\lambda_\Psi}{2}|\Psi_I|^4\right)
 &=\sqrt{-g}\,X\left(3m_\Psi^2f^2+
                         \frac{3\lambda_\Psi}{2}f^4\right),\\
 \sqrt{-\widehat g}\,\frac{42}{L_8^2}
 &=\sqrt{-g}\,\frac{42X}{L_8^2}.
 \end{aligned}
 \label{eq:neutral_reduction_terms}
\end{equation}
The nonminimal charged curvature coupling is reduced by multiplying
\eqref{eq:Ricci} by $\mathcal F$ from \eqref{eq:F_definition}
and integrating the $\Box_5\phi$ term by parts,
\begin{equation}
 \begin{aligned}
 \sqrt{-\widehat g}\,\mathcal F R_8
 &=\sqrt{-g}\,\mathcal F
 \left(R_5-\frac12(\partial\phi)^2
             -\frac1{\sqrt3}\Box_5\phi\right)\\
 &\simeq\sqrt{-g}\left[
 \mathcal F R_5-\frac{\mathcal F}{2}(\partial\phi)^2
 +\frac1{\sqrt3}\partial_\mu\mathcal F\,\partial^\mu\phi\right].
 \end{aligned}
 \label{eq:curvature_reduction_steps}
\end{equation}
Here $\simeq$ denotes equality up to a boundary derivative.
The sum of \eqref{eq:reduced_densities}--\eqref{eq:curvature_reduction_steps}
is the five-dimensional action
\begin{align}
 S_5=\frac1{2\kappa_5^2}\int\dd^5x\sqrt{-g}\,\Big[
 &\mathcal F R_5-\frac{\mathcal F}2(\partial\phi)^2-3(\partial f)^2
 +\frac1{\sqrt3}\partial\mathcal F\cdot\partial\phi-V(X,f) \nonumber\\
 &-\frac{\eps}{4X}G^a_{\mu\nu}G^{a\mu\nu}
 -\eps(D_\mu\boldsymbol\Phi)^\dagger D^\mu\boldsymbol\Phi
 -\eps X(m_0^2+3\eta f^2)\boldsymbol\Phi^\dagger\boldsymbol\Phi
 -\eps\lambda_8 X(\boldsymbol\Phi^\dagger\boldsymbol\Phi)^2\Big], \label{eq:S5}\\
 V(X,f)&=X\,U(f)+3k^2X^2f^2,\qquad
 U(f)=-\frac{42}{L_8^2}+3m_\Psi^2f^2+\frac{3\lambda_\Psi}{2}f^4 .
 \label{eq:potential}
\end{align}
Here $\kappa_5^2=\kappa_8^2/\operatorname{Vol}(T^3)$, where
$\operatorname{Vol}(T^3)=(2\pi/|k|)^3$. The terms $XU(f)$ and
$3k^2X^2f^2$ follow respectively from the neutral potential
and the internal derivatives in \eqref{eq:neutral_reduction_terms}.
At $\boldsymbol\Phi=0$ the scalar kinetic metric is
$\dd\phi^2+6\,\dd f^2$.

In the neutral probe limit, $\mathcal F=1$ and its derivative
coupling vanishes. The curvature contribution to the charged
scalar mass can then be obtained directly from the first line
of \eqref{eq:curvature_reduction_steps},
\begin{equation}
 \sqrt{-\widehat g}\,\xi R_8
 \boldsymbol\Phi^\dagger\boldsymbol\Phi
 =\sqrt{-g}\,\xi
 \left[R_5-\frac12(\partial\phi)^2
 -\frac1{\sqrt3}\Box_5\phi\right]
 \boldsymbol\Phi^\dagger\boldsymbol\Phi .
 \label{eq:curvature_mass_reduction}
\end{equation}
To fix the meaning of the mass used below, expand the charged
five-dimensional Lagrangian to quadratic order in the doublet on a
neutral background $(g_{\mu\nu},\phi,f)$. We define
$m_\Phi^2$ as the coefficient of
$-\eps\boldsymbol\Phi^\dagger\boldsymbol\Phi$ in this
\emph{five-dimensional quadratic fluctuation operator}, after
the eight-dimensional curvature term has been reduced. It is
therefore a background-dependent effective mass, distinct from
the parent parameter $m_0^2$. In this notation the quadratic
charged action reads
\begin{equation}
 S_5^{(2)}[\boldsymbol\Phi]
 =-\frac{\eps}{2\kappa_5^2}\int\dd^5x\sqrt{-g}\,
 \left[(D_\mu\boldsymbol\Phi)^\dagger D^\mu\boldsymbol\Phi
 +m_\Phi^2\boldsymbol\Phi^\dagger\boldsymbol\Phi\right],
 \label{eq:effective_mass_definition}
\end{equation}
where the superscript denotes quadratic order in $\boldsymbol\Phi$
and the background metric and gauge field are held fixed.
Reading the Yang--Mills and doublet coefficients from
\eqref{eq:S5} and \eqref{eq:curvature_mass_reduction} then gives
\begin{equation}
 \boxed{\;\begin{gathered}
 Z(\phi)=X^{-1},\qquad \lambda_\Phi(\phi)=\lambda_8X,\qquad
 Z(\phi)\lambda_\Phi(\phi)=\lambda_8,\\
 m_\Phi^2(\phi,f)=X(m_0^2+3\eta f^2)
 +\xi\left[R_5-\frac12(\partial\phi)^2
 -\frac{1}{\sqrt3}\Box_5\phi\right].
 \end{gathered}\;}
 \label{eq:tetracritical_Z}
\end{equation}
In the probe limit the $m_\Phi^2$ entry is evaluated on a neutral
background. At finite charged backreaction the unspecialized
$\mathcal F R_5$ and derivative mixing in \eqref{eq:S5} supply the
metric variation of the same curvature operator.

At the product vacuum $\AdS_5\times T^3$ the breathing mode and the
winding amplitude are constant. We normalize $X=1$ there and write
$f_{\rm UV}$ for the winding amplitude. Such a background is a vacuum
only when the scalar equations have no constant-field source and the
Einstein equations admit the required $\AdS_5$ curvature. The first
requirement is $\partial_X V=\partial_fV=0$. The second is
$V=-12/L_5^2$. The two stationarity
equations follow explicitly from \eqref{eq:potential},
\begin{align}
 \frac{1}{6f}\frac{\partial V}{\partial f}
 &=X\left(m_\Psi^2+\lambda_\Psi f^2+k^2X\right),&
 \frac{\partial V}{\partial X}
 &=U(f)+6k^2Xf^2.
 \label{eq:vacuum_stationarity_steps}
\end{align}
For a nonzero winding amplitude $f_{\rm UV}$, the first equation
at $X=1$ gives
\begin{equation}
 m_\Psi^2=-k^2-\lambda_\Psi f_{\rm UV}^2.
 \label{eq:neutral_mass_stationarity}
\end{equation}
Substitute this into $U(f_{\rm UV})$ and then impose the second
stationarity equation. The two steps are
\begin{align}
 U(f_{\rm UV})
 &=-\frac{42}{L_8^2}-3k^2f_{\rm UV}^2
   -\frac{3\lambda_\Psi}{2}f_{\rm UV}^4,
 \nonumber\\
 0=\left.\partial_XV\right|_{1,f_{\rm UV}}
 &=U(f_{\rm UV})+6k^2f_{\rm UV}^2
 \quad\Longrightarrow\quad
 \frac{42}{L_8^2}=3k^2f_{\rm UV}^2
 -\frac{3\lambda_\Psi}{2}f_{\rm UV}^4.
 \label{eq:neutral_cosmological_stationarity}
\end{align}
Finally, the value of the potential at the stationary point is
\begin{equation}
 V(1,f_{\rm UV})=U(f_{\rm UV})+3k^2f_{\rm UV}^2
 =-3k^2f_{\rm UV}^2
 =-\frac{12}{L_5^2}.
 \label{eq:external_vacuum_radius}
\end{equation}
We set $L_5=1$ after this equality. The vacuum equations thus fix
the three parent parameter combinations
\begin{align}
m_\Psi^2+k^2+\lambda_\Psi f_{\rm UV}^2&=0,&
\frac{42}{L_8^2}&=3k^2f_{\rm UV}^2
-\frac32\lambda_\Psi f_{\rm UV}^4,&
k^2f_{\rm UV}^2&=4.
\label{eq:vacuum_relations}
\end{align}
These relations solve the neutral field equations and the independent
external and internal Einstein equations at the product vacuum.
On this vacuum $R_5=-20$, so the
doublet mass in Eq.~\eqref{eq:tetracritical_Z} equals
$m_0^2+3\eta f_{\rm UV}^2-20\xi$. The source-free ultraviolet
Breitenlohner--Freedman (BF) condition therefore fixes one charged
mass parameter,
\begin{align}
 m_0^2-20\xi+3\eta f_{\rm UV}^2&=m_{\rm UV}^2=-4,&
 \lambda_8&>0,\qquad\eps>0.
 \label{eq:parameters}
\end{align}
The curvature and portal couplings, the winding parameters and the
neutral source are held fixed in any thermodynamic comparison.

Both neutral profiles are dynamical throughout the flow. In particular,
substitution of $f=f_{\rm UV}$ into the neutral scalar equation
\eqref{eq:parent_scalar_equations} gives a nonzero radial source
whenever $X\ne1$,
\begin{equation}
 \left.\partial_fV(X,f)\right|_{f=f_{\rm UV}}
 =6k^2f_{\rm UV}X(X-1).
 \label{eq:obstruction_constant_winding}
\end{equation}
The infrared limit $f\to0$ is therefore an asymptotic property of
the two-field interpolation.

It remains to verify that fields excluded by the reduction have
vanishing equations of motion. The gradients of the three windings
have diagonal internal stress and no external--internal cross term,
\begin{equation}
 \sum_I\partial_{y^J}\Psi_I^*\partial_{y^K}\Psi_I
 =k^2f^2\delta_{JK},\qquad
 2\operatorname{Re}\sum_I
 \partial_\mu\Psi_I^*\partial_{y^J}\Psi_I=0.
 \label{eq:internal_stress_consistency}
\end{equation}
The phase equation of each $\Psi_I=f\exp(iky^I)$ is likewise
identically satisfied,
\begin{equation}
 \partial_M\!\left(\sqrt{-\widehat g}\,f^2
 \widehat g^{MN}\partial_N(ky^I)\right)
 =\partial_{y^I}\!\left(kX^2\sqrt{-g}\,f^2\right)=0.
 \label{eq:neutral_phase_consistency}
\end{equation}
The omitted internal Yang--Mills equation vanishes separately,
since no internal field strength or doublet covariant derivative
remains,
\begin{equation}
 G_{\mu I}^a=G_{IJ}^a=0,\qquad
 D_I\boldsymbol\Phi=0
 \quad\Longrightarrow\quad
 \left.\frac{\delta S_8}{\delta B_I^a}\right|_{B_I=0}=0.
 \label{eq:internal_gauge_consistency}
\end{equation}
For the nonminimal Einstein term the only internal second derivative
is proportional to the internal metric,
\begin{equation}
 \nabla_\mu\nabla_I\mathcal F=0,\qquad
 \nabla_I\nabla_J\mathcal F
 =-\frac{\widehat g_{IJ}}{2X^2}
 \partial_\mu X\,\partial^\mu\mathcal F.
 \label{eq:internal_curvature_consistency}
\end{equation}
Inserting \eqref{eq:internal_stress_consistency} and
\eqref{eq:internal_curvature_consistency} into the mixed and
traceless internal projections of the Einstein equation
\eqref{eq:8dEinstein} gives
\begin{equation}
 \frac{\delta S_8}{\delta\widehat g_{\mu I}}=0,
 \qquad
 \frac{\delta S_8}{\delta\widehat g_{IJ}}
 -\frac13\widehat g^{IJ}\widehat g_{KL}
 \frac{\delta S_8}{\delta\widehat g_{KL}}=0.
 \label{eq:omitted_equations}
\end{equation}
The common internal trace and the external Einstein equations
are the variations of \eqref{eq:S5} with respect to $\phi$
and $g_{\mu\nu}$. Equations \eqref{eq:internal_gauge_consistency}
and \eqref{eq:omitted_equations} thus complete the background
uplift check.

For the ordered ansatz
$\boldsymbol\Phi=(0,s)^T/\sqrt2$, $B_t^3=b$ and $B_{x_1}^1=w$,
the nonzero gauge strengths and scalar bilinears read
\begin{align}
 G^3_{rt}&=b',&G^1_{rx_1}&=w',&G^2_{tx_1}&=-bw,
 \nonumber\\
 (D_\mu\boldsymbol\Phi)^\dagger D^\mu\boldsymbol\Phi
 &=\frac12 g^{rr}(s')^2
   +\frac{s^2}{8}\bigl(g^{tt}b^2+g^{x_1x_1}w^2\bigr),
 \nonumber\\
 (D_t\boldsymbol\Phi)^\dagger D_{x_1}\boldsymbol\Phi
 +\mathrm{c.c.}
 &=bw\,\boldsymbol\Phi^\dagger
    \{\tau^3,\tau^1\}\boldsymbol\Phi=0,&
 \left.\frac{\partial L_{\rm ch}}{\partial B_\mu^a}\right|_{\boldsymbol\Phi}
 &=-\frac{s^2}{4}B^{a\mu}.
 \label{eq:ordered_consistency}
\end{align}
Since $\{\tau^3,\tau^1\}=0$ and
$(\tau^1)^2=(\tau^3)^2=\mathbf1/4$, the off-diagonal scalar stress
and the equation of the unused doublet component vanish. The Yang--Mills
stress $G_{tM}^aG_{x_1}{}^{Ma}$ also vanishes for the three distinct
adjoint components displayed above. Fluctuations beyond these retained
background fields are considered in Section~\ref{sec:parent_checks}.

The kinetic terms also admit a direct action-level check. With
$\boldsymbol\Phi=(0,s)^T/\sqrt2$, the curvature factor becomes
$\mathcal F=1-\eps\xi s^2/2$. The $(\phi,s)$ kinetic terms in
\eqref{eq:S5} are
\begin{equation}
 -\frac12\left[
 \mathcal F(\partial\phi)^2+
 \frac{2\eps\xi s}{\sqrt3}\partial_\mu\phi\,\partial^\mu s
 +\eps(\partial s)^2\right].
 \label{eq:kinetic_mixing_explicit}
\end{equation}
The five-dimensional Weyl transformation to a constant Einstein
coefficient then gives the determinant of the scalar kinetic matrix,
\begin{equation}
 \frac{\eps}{\mathcal F}
 +\frac{7\eps^2\xi^2s^2}{3\mathcal F^2}>0
 \qquad(\mathcal F>0,\ \eps>0).
 \label{eq:kinetic}
\end{equation}
Its first principal minor is positive under the same conditions.
The independent winding-scalar and Yang--Mills kinetic coefficients
are also positive. Thus $\mathcal F>0$ specifies the regular
positive-Einstein-coefficient region of the reduced theory.
The reduction is consistent for external anisotropic fields within
the retained ansatz, while the coupled Kaluza--Klein fluctuation
spectrum is addressed separately in Section~\ref{sec:parent_checks}.

\subsection{Ultraviolet and infrared scaling geometries}\label{sec:background}
The exact UV solution has $X=1$, $f=f_{\rm UV}$, and geometry
$\AdS_5\times T^3$. For the analytic conformal reference of
Section~\ref{sec:perturbation} we use unit horizon radius and
\begin{equation}
 ds_0^2=u^2\left[-f_u(u)\,\dd t^2+\dd\mathbf x_3^2\right]
 +\frac{\dd u^2}{u^2f_u(u)},\qquad f_u(u)=1-u^{-4},
 \label{eq:AdS_background}
\end{equation}
so that $T=1/\pi$ in that reference normalization.

We choose to drive the ultraviolet theory with one relevant neutral
operator. This is a choice of the family of flows to be studied, not
a condition imposed by dimensional reduction. Allowing two relevant
neutral sources would require specifying and varying both of them.
The two neutral fluctuations retained by the reduction are
$\delta\phi$ and $\sqrt6\,\delta f$, both canonically normalized.
Because they mix, the relevant and irrelevant operators correspond
to the eigenvectors of their mass matrix, not to $\phi$ and $f$
separately.
Expanding \eqref{eq:potential} to quadratic order at
$X=1$, $f=f_{\rm UV}$ gives
\begin{equation}
 V=V_{\rm UV}+\frac12
 \begin{pmatrix}\delta\phi&\sqrt6\,\delta f\end{pmatrix}
 \begin{pmatrix}
 2k^2 f_{\rm UV}^2&\sqrt2\,k^2 f_{\rm UV}\\
 \sqrt2\,k^2 f_{\rm UV}&2\lambda_\Psi f_{\rm UV}^2
 \end{pmatrix}
 \begin{pmatrix}\delta\phi\\\sqrt6\,\delta f\end{pmatrix}
 +O\bigl((|\delta\phi|+|\delta f|)^3\bigr).
 \label{eq:BF_matrix_test}
\end{equation}
The two squared masses are the eigenvalues of the displayed
$2\times2$ quadratic form. Their product is
\begin{equation}
 m_+^2m_-^2
 =2k^2f_{\rm UV}^2
 \bigl(2\lambda_\Psi f_{\rm UV}^2-k^2\bigr).
 \label{eq:relevant_neutral_criterion}
\end{equation}
Diagonalizing the same quadratic form gives
\begin{equation}
 m_\pm^2=f_{\rm UV}\Big[(k^2+\lambda_\Psi)f_{\rm UV}
 \pm\sqrt{(k^2-\lambda_\Psi)^2f_{\rm UV}^2+2k^4}\,\Big].
 \label{eq:singlet_masses}
\end{equation}
When $2\lambda_\Psi f_{\rm UV}^2<k^2$, their product is
negative and thus $m_-^2<0<m_+^2$. In five-dimensional AdS
units the BF bound is $m_-^2L_5^2\ge-4$. Since
$k^2f_{\rm UV}^2=4$ and $L_5=1$ by
\eqref{eq:vacuum_relations}, the shifted product of the same two
eigenvalues is
\begin{equation}
 (m_+^2+4)(m_-^2+4)
 =8\bigl(6+3\lambda_\Psi f_{\rm UV}^2-k^2\bigr).
 \label{eq:shifted_BF_product}
\end{equation}
Here $m_+^2>0$, so the strict BF condition $m_-^2>-4$ is
equivalent to positivity of \eqref{eq:shifted_BF_product}.
Combining it with the negative-product condition
\eqref{eq:relevant_neutral_criterion} gives the range in which
one neutral mode is relevant and the other irrelevant,
\begin{equation}
 2\lambda_\Psi f_{\rm UV}^2<k^2<6+3\lambda_\Psi f_{\rm UV}^2 .
 \label{eq:singlet_window}
\end{equation}
In standard $\AdS_5$ quantization, the mode with
$-4<m_-^2L_5^2<0$ is dual to an operator of dimension between
two and four, so its source is relevant. The mode with $m_+^2>0$
has dimension greater than four. Its irrelevant source is set to
zero. Thus \eqref{eq:singlet_window} leaves one neutral deformation
$J$ to be held fixed when the charged phases are compared.
The independent internal shape and relative-winding sectors require
their own mass matrices and are not covered by this singlet test.

To identify an infrared endpoint we choose the locally $\AdS_8$
branch of the same parent theory. Setting the winding amplitude to
zero removes its internal gradient energy and permits the torus
directions to acquire the same local warp factor as the external
directions. This is a property of the selected infrared solution,
not a consequence of writing the infrared coordinate as $r=0$.
At $f=0$ the same eight-dimensional theory has an $\AdS_8$ solution
with radius $L_8$ fixed by its cosmological constant.
Let $r$ be its Poincar\'e radial coordinate, with the horizon at
$r=0$. Since $\widehat g_{rr}=Xg_{rr}=L_8^2/r^2$ and the internal
metric is $X^{-1}=r^2/L_8^2$ up to a constant rescaling of $y^I$, the
same $r$ is the proper radial coordinate of the five-dimensional
metric \eqref{eq:neutral_radial_metric}. Comparison with
Eq.~\eqref{eq:uplift} then gives the zero-temperature infrared
solution
\begin{equation}
 X=\frac{L_8^2}{r^2},\qquad e^{A}=c_A\frac{r^2}{L_8^2},\qquad
 f=r^{-3}e^{-|k|L_8^2/r}
 \left[f_{\rm IR}+O(r)\right],
 \label{eq:IR}
\end{equation}
where the constant $c_A$ relates the infrared and ultraviolet
normalizations of the boundary coordinates. The metric and breathing
mode receive corrections of order $f^2$.
The power in \eqref{eq:IR} is fixed by the winding equation itself.
In the locally $\AdS_8$ coordinate used here its leading terms are
\begin{equation}
 r^2f''+8rf'-\left(\frac{k^2L_8^4}{r^2}
 +m_\Psi^2L_8^2\right)f=O(f^3),
 \label{eq:IR_winding_ode}
\end{equation}
and its entire linear solution is expressed in modified Bessel
functions. Set $x=|k|L_8^2/r$. The substitution
$f=r^{-7/2}g(x)$ gives, with $D=8$ the parent bulk dimension,
\begin{equation}
 x^2g''+xg'-(x^2+\nu^2)g=0,
 \qquad \nu^2=\frac{(D-1)^2}{4}+m_\Psi^2L_8^2,
 \label{eq:winding_bessel_index}
\end{equation}
whose independent solutions are $K_\nu(x)$ and $I_\nu(x)$.
At the Poincar\'e horizon $r\to0$ the $I_\nu$ branch grows
exponentially and is excluded by regularity. The regular infrared
winding is therefore
\begin{equation}
 \boxed{\;f(r)=c_K\,r^{-7/2}K_\nu\!\left(\frac{|k|L_8^2}{r}\right)+O(f^3).\;}
 \label{eq:winding_Bessel}
\end{equation}
Its large-argument expansion
$K_\nu(x)=\sqrt{\pi/(2x)}\,e^{-x}[1+O(x^{-1})]$ reproduces the
decay $r^{-3}e^{-|k|L_8^2/r}$ stated in \eqref{eq:IR}. The index is real provided the neutral parent
mass obeys the usual $\AdS_8$ scalar BF condition
$m_\Psi^2L_8^2\ge-(D-1)^2/4$. The winding perturbation is then
exponentially suppressed compared with any algebraic correction.
It supplies the infrared boundary condition of the coupled
interpolation, whose single constant $c_K$ is fixed by matching to the
ultraviolet. The torus remains a compact quotient.
the local isometry, not the spacetime dimension, changes along
the flow.

Figure~\ref{fig:rgconditions} summarizes how the same parent action
determines the geometry, the charged instability problems and the
thermodynamic comparison. The horizontal geometric arrows depict
the selected ultraviolet-to-infrared branch. Its existence for
arbitrary ultraviolet source data is not assumed. The lower row
also distinguishes endpoint calculations from the additional
requirements at a complete-flow common onset.
\begin{figure}[htbp]
 \centering
 \begin{tikzpicture}[x=1mm,y=1mm,>=Latex,
   font=\sffamily\footnotesize,
   parent/.style={rounded corners=2.5mm,draw=blue!60!black,
     line width=.9pt,fill=blue!78!black,text=white,
     text width=138mm,minimum height=12mm,align=center,
     inner xsep=2mm,inner ysep=1.5mm},
   geo/.style={rounded corners=2mm,draw=#1!72!black,
     line width=.85pt,fill=#1!7,
     text width=42.5mm,minimum height=22mm,align=center,
     inner xsep=2mm,inner ysep=1.5mm},
   probe/.style={rounded corners=2mm,draw=#1!65!black,
     line width=.65pt,fill=#1!4,
     text width=42.5mm,minimum height=16mm,align=center,
     inner xsep=2mm,inner ysep=1.5mm},
   arr/.style={-{Latex[length=2.5mm,width=1.8mm]},
     line width=1.05pt,draw=blue!72!black},
   downarr/.style={-{Latex[length=2.1mm,width=1.5mm]},
     line width=.75pt,draw=black!49}]
  \fill[gray!5,rounded corners=2.2mm]
    (-79,37) rectangle (79,73);
  \fill[gray!4,rounded corners=2.2mm]
    (-79,0) rectangle (79,36);
  \node[parent] (parent) at (0,82) {
    {\bfseries Eight-dimensional Einstein--matter theory}\\[1mm]
    Fixed couplings};
  \node[font=\sffamily\bfseries\scriptsize,text=blue!70!black,
    anchor=west] at (-76,69.5) {GEOMETRY AND RADIAL RG FLOW};
  \node[geo=blue] (uv) at (-54,52) {
    {\bfseries Ultraviolet}\\[1mm]
    $\AdS_5\times T^3$\\
    Scalar winding};
  \node[geo=orange] (cross) at (0,52) {
    {\bfseries Neutral RG flow}\\[1mm]
    Fixed source\\
    Running torus volume};
  \node[geo=teal] (ir) at (54,52) {
    {\bfseries Infrared}\\[1mm]
    Locally $\AdS_8$\\
    Hyperscaling violation};
  \draw[arr] (uv.east)--(cross.west);
  \draw[arr] (cross.east)--(ir.west);
  \draw[downarr] (parent.south)--(cross.north)
    node[pos=.25,right,font=\scriptsize,text=black!65]
      {Three-torus reduction};
  \node[font=\sffamily\bfseries\scriptsize,text=blue!70!black,
    anchor=west] at (-76,32.5) {CHARGED MODES};
  \node[probe=blue] (uvmode) at (-54,15) {
    {\bfseries UV}\\[1mm]
    Zero modes};
  \node[probe=orange] (fullmode) at (0,15) {
    {\bfseries Full flow}\\[1mm]
    Common onset};
  \node[probe=teal] (irmode) at (54,15) {
    {\bfseries IR}\\[1mm]
    Spectra};
  \draw[downarr] (cross.south)--(fullmode.north);
  \draw[->,dashed,draw=black!48,line width=.7pt]
    (uvmode.east)--(fullmode.west);
  \draw[->,dashed,draw=black!48,line width=.7pt]
    (irmode.west)--(fullmode.east);
  \node[rounded corners=2mm,draw=purple!70!black,
    fill=purple!5,line width=.9pt,text width=143mm,
    minimum height=14mm,align=center,inner sep=2mm] (thermo)
    at (0,-15) {
      {\bfseries Phase competition}\\[1mm]
      Scalar \enspace\textbullet\enspace
      Vector \enspace\textbullet\enspace Coexistence};
  \draw[downarr] (fullmode.south)--(thermo.north);
  \node[rounded corners=1.5mm,draw=gray!55,
     fill=gray!8,text width=140mm,minimum height=12mm,
     align=center,inner sep=1.5mm] (ensembles) at (0,-38) {
       {\bfseries Ensembles}\\[1mm]
       Fixed potential \enspace\textbullet\enspace Fixed density};
  \draw[downarr] (thermo.south)--(ensembles.north);
 \end{tikzpicture}
 \caption{Geometry and charged-order analysis in the fixed
  eight-dimensional theory. The upper row shows the selected
  ultraviolet-to-infrared flow. The lower row uses exact UV zero
  modes and IR spectra to constrain a common onset on the full
  geometry, where both charged sources must vanish and the onset
  curves must cross. Electric response and quartic interactions
  determine the free-energy comparison of scalar, vector and
  coexisting orders. Coexistence requires a locally stable mixed
  minimum with both condensates nonzero. The two ensembles fix
  either the chemical potential or the charge density and are
  related by a Legendre transform.}
 \label{fig:rgconditions}
\end{figure}

At the infrared endpoint the homogeneous charged scalar has
\begin{equation}
 M_{\rm IR}^2L_8^2=L_8^2m_0^2-56\xi,
 \qquad M_{\rm IR}^2L_8^2\ge-\frac{(D-1)^2}{4},
 \label{eq:charged_IR_BF}
\end{equation}
where the last inequality is its $\AdS_8$ BF condition. After
Eq.~\eqref{eq:parameters} is imposed it constrains the remaining
combination of $\xi$ and $\eta$.

In the eight-dimensional metric the external and internal
directions of Eq.~\eqref{eq:IR} share the local scale $r^2/L_8^2$.
The five-dimensional Einstein frame therefore realizes $z=1$ and
$\theta=-3$, the case $n=3$ of Eq.~\eqref{eq:general_theta}, where
$n$ counts the compact internal directions of $T^n$ and hence
equals three for the parent ansatz \eqref{eq:uplift}. The
torus identifications are retained, so the scaling region is a
locally higher-dimensional part of the same bulk.

For a local $\AdS_{d+n}$ geometry reduced on $n$ torus cycles,
the internal horizon volume supplies an additional factor of
the temperature. With $d_s=d-2$ external spatial directions,
the five-dimensional hyperscaling notation generalizes to
\begin{equation}
 z=1,\qquad\theta=-n,\qquad
 s_{\rm th}\propto T^{(d_s-\theta)/z}=T^{d_s+n}.
 \label{eq:general_theta}
\end{equation}
The boundary field theory retains its original $d_s+1$
dimensions, while the parent metric determines the exponent
through the internal horizon area.

The neutral uniform thermal family and its probe observables have the scaling limits
\begin{center}
\begin{tabular}{lll}
\toprule
Observable & UV $\AdS_5$ & IR locally $\AdS_8$\\
\midrule
Entropy density & $s_{\rm th}\propto T^3$ & $s_{\rm th}\propto T^6$\\
Zero-density optical scaling & $\operatorname{Re}\sigma\propto\omega$ & $\operatorname{Re}\sigma\propto\omega^4$\\
\bottomrule
\end{tabular}
\end{center}
These are limits of the neutral flow on which the charged
orders of Section~\ref{sec:charged} are defined.
For a zero-density transverse gauge perturbation
$a_x(r)e^{-i\omega t}$, the local $\AdS_8$ Maxwell equation gives
\begin{equation}
 \partial_r\big(r^6\partial_r a_x\big)
 +\omega^2 L_8^4 r^2a_x=0,
 \qquad \operatorname{Re}\sigma(\omega)\propto\omega^4.
 \label{eq:IR_optical_equation}
\end{equation}
This is the external component of
$\partial_M(\sqrt{-\widehat g}\,G^{Mx})=0$ in the local infrared
metric. The crossover functions and finite coefficients depend
on the complete neutral background.

\subsection{Radial evolution and holographic boundary data}
In the absence of the charged orders the reduced theory
contains two independent scalar profiles.
For an isotropic black-brane metric, write
\begin{equation}
 \dd s_5^2=\frac{\dd r^2}{h(r)}
 +e^{2A(r)}\big[-h(r)\dd t^2+\dd\mathbf x_3^2\big],
 \quad \phi=\phi(r),\quad f=f(r).
 \label{eq:neutral_radial_metric}
\end{equation}
The function $A(r)$ is the warp factor along the noncompact directions,
and $h(r)$ is the blackening function.
At zero charged backreaction the two scalar equations follow
directly from the action-level potential \eqref{eq:potential}.
\begin{align}
 \phi''+\left(4A'+\frac{h'}h\right)\phi'
 &=\frac{\partial_\phi V}{h},&
 f''+\left(4A'+\frac{h'}h\right)f'
 &=\frac{\partial_f V}{6h}.
 \label{eq:neutral_scalars}
\end{align}
The two derivatives of the reduced potential entering these equations
are
\begin{equation}
 \partial_\phi V=\frac{XU(f)+6k^2X^2f^2}{\sqrt3},\qquad
 \partial_fV=6Xf\left(m_\Psi^2+\lambda_\Psi f^2+k^2X\right).
 \label{eq:neutral_explicit_forces}
\end{equation}
The independent Einstein evolution equations and the radial
constraint can be written
\begin{align}
 A''&=-\frac16\big[(\phi')^2+6(f')^2\big],&
 h''+4A'h'&=0,\nonumber\\
 6A'h'&+h\big[24(A')^2-(\phi')^2-6(f')^2\big]+2V=0.
 \label{eq:neutral_einstein}
\end{align}
The first integral $h'=c_he^{-4A}$ is the uncharged,
isotropic blackening relation. With the ordinary horizon and
boundary normalizations it gives
\begin{equation}
 h(r)=\frac{\displaystyle\int_{r_h}^r e^{-4A(v)}\,\dd v}
 {\displaystyle\int_{r_h}^\infty e^{-4A(v)}\,\dd v}.
 \label{eq:neutral_blackening}
\end{equation}
At zero temperature the neutral equations admit a local first-order
description. Wherever the flow can be expressed as a gradient flow
in the two-field space, a function $W(\phi,f)$ satisfies
\begin{align}
 V&=\frac12(\partial_\phi W)^2
 +\frac1{12}(\partial_fW)^2-\frac13W^2,
 \label{eq:twofield_W}\\
 A'&=-\frac W6,\qquad
 \phi'=\partial_\phi W,\qquad
 f'=\frac16\partial_fW.
 \label{eq:twofield_flow}
\end{align}
The potential is the one derived in \eqref{eq:potential}, with no
additional scalar function introduced as a coupling. At the product
vacuum $W=-6/L_5$ with vanishing field derivatives. At $f=0$ the
infrared branch is explicitly
\begin{equation}
 W_{\rm IR}(\phi)=-\frac{12}{L_8}\sqrt X,
 \qquad V(X,0)=-\frac{42}{L_8^2}X.
 \label{eq:IR_W}
\end{equation}
Substituting \eqref{eq:IR_W} into \eqref{eq:twofield_W}
reproduces the potential exactly on that boundary of field space.
The winding term $3k^2X^2f^2$ prevents this one-field expression
from solving the full two-field problem at finite $f$.
Consequently the endpoint formulas are exact but the interpolating
solution is specified by the coupled equations
\eqref{eq:neutral_scalars}--\eqref{eq:neutral_einstein}, rather than
by an assumed elementary superpotential.

For a finite electric field or charged condensate the $O(\eps)$
stress modifies the radial equations. Thus
\eqref{eq:neutral_scalars}--\eqref{eq:neutral_blackening} specify
the uncharged background. The charged stationary solutions use
\eqref{eq:backreacted_ansatz} below and the complete action
\eqref{eq:S5}.

The ultraviolet expansion follows from the neutral quadratic
form \eqref{eq:BF_matrix_test}. In the window
\eqref{eq:singlet_window}, the negative squared mass $m_-^2$
of \eqref{eq:singlet_masses} corresponds to a relevant
boundary operator of scaling dimension
\begin{equation}
 \Delta_J=2+\sqrt{4+m_-^2L_5^2},\qquad
 2<\Delta_J<4.
 \label{eq:source_dimension}
\end{equation}
We denote by $e_-$ the unit eigenvector of this explicitly
displayed quadratic form corresponding to $m_-^2$, with its
overall sign fixed once and for all.
The source $J$ is the coefficient of the asymptotic mode
$e^{-(4-\Delta_J)r}e_-$ in the normalization $A(r)\sim r$.
The source of the second, irrelevant eigenmode is set to zero.

For thermal states and competing charged orders we extend the
radial metric to allow an anisotropic spatial direction,
\begin{equation}
 ds_5^2=\frac{\dd r^2}{h}
 +e^{2A}\big[-h\,\dd t^2+e^{2B}\dd x_1^2+\dd x_2^2+\dd x_3^2\big],
 \qquad (B_t^3,B_{x_1}^1,\boldsymbol\Phi)=\Big(b,w,\frac{(0,s)^T}{\sqrt2}\Big).
 \label{eq:backreacted_ansatz}
\end{equation}
The function $B(r)$ measures the anisotropy along $x_1$ relative
to the $x_2,x_3$ plane.
At the horizon $r=r_h$ we impose $h(r_h)=b(r_h)=0$. In units
with $2\kappa_5^2=1$ the temperature, Wald entropy and Cartan density are
\begin{equation}
 T=\frac{e^{A_h}h'_h}{4\pi},\qquad
 s_{\rm th}=4\pi\mathcal F_h e^{3A_h+B_h},\qquad
 \rho=\eps\lim_{r\to\infty}\frac{e^{2A+B}}X b' .
 \label{eq:thermo}
\end{equation}
The gauge-field canonical momenta computed from \eqref{eq:S5}
are
\begin{equation}
 \begin{aligned}
 \Pi_b&=-\eps\sqrt{-g}\,X^{-1}G^{3rt}
 =\eps\frac{e^{2A+B}}X b',\\
 \Pi_w&=-\eps\sqrt{-g}\,X^{-1}G^{1rx_1}
 =-\eps\frac{h e^{2A-B}}X w'.
 \end{aligned}
 \label{eq:boundary_momenta}
\end{equation}
At $A(r)\sim r$, $B(r)\to0$ and $X(r)\to1$, their asymptotic
coefficients and the neutral source are fixed by
\begin{align}
 \begin{pmatrix}\phi\\\sqrt6(f-f_{\rm UV})\end{pmatrix}
 &=J e_-e^{-(4-\Delta_J)r}+\cdots,&
 J&=\lim_{r\to\infty}e^{(4-\Delta_J)r}
 e_-\cdot\begin{pmatrix}\phi\\\sqrt6(f-f_{\rm UV})\end{pmatrix},
 \label{eq:J}\\
 b(r)&=\mu-\frac{\rho}{2\eps}e^{-2r}+\cdots,&
 w(r)&=w_{(s)}+\frac{\langle J_1^{\,x_1}\rangle}{2\eps}e^{-2r}+\cdots,
 \nonumber\\
 s(r)&=e^{-2r}\big(s_{\rm src}r+s_{\rm vev}+\cdots\big),&
 \rho&=\lim_{r\to\infty}\Pi_b,\quad
 \langle J_1^{\,x_1}\rangle=\lim_{r\to\infty}\Pi_w.
 \label{eq:charged_falloffs}
\end{align}
The $s$ expansion uses the saturated $\AdS_5$ BF mass in
\eqref{eq:parameters}. We impose
$s_{\rm src}=w_{(s)}=0$, remove the irrelevant neutral source and
hold $J$ fixed. The renormalized Dirichlet prescription follows
Ref.~\cite{Skenderis:2002wp}.

The ultraviolet four-dimensional sources consist of the boundary
metric $\gamma_{(0)ij}$, the neutral deformation $J$, the gauge
field $B_{(0)i}^a$ and the doublet source
$\boldsymbol\Phi_{(s)}$. Their renormalized variation is
\begin{equation}
 \delta S_{\rm ren}
 =\int\dd^4x\sqrt{-\gamma_{(0)}}\left(
 \frac12\langle T^{ij}\rangle\delta\gamma_{(0)ij}
 +\langle\mathcal O_J\rangle\delta J
 +\langle J_a^{\,i}\rangle\delta B_{(0)i}^{a}
 +\langle\boldsymbol{\mathcal O}_\Phi^\dagger\rangle
           \delta\boldsymbol\Phi_{(s)}
 +\mathrm{c.c.}\right).
 \label{eq:full_dictionary}
\end{equation}
The generalized Gibbons--Hawking term for $\mathcal F R_5$ and
local counterterms define $S_{\rm ren}$. Its gauge variation
gives the Ward identity in \eqref{eq:Ward_identity}.

At fixed $J$, $\boldsymbol\Phi_{(s)}$ and $w_{(s)}$, the thermodynamic
variations in the two ensembles are
\begin{equation}
 \begin{aligned}
 \dd\Omega&=-s_{\rm th}\,\dd T-\rho\,\dd\mu,&
 F&=\Omega+\mu\rho,&
 \dd F&=-s_{\rm th}\,\dd T+\mu\,\dd\rho .
 \end{aligned}
 \label{eq:general_first_law}
\end{equation}
At common boundary sources the relevant stationary branches are
\begin{equation}
 \begin{array}{c|cccc}
 &\text{normal}&s\text{-wave}&p\text{-wave}&s+p\\ \hline
 s&0&\ne0&0&\ne0\\
 w&0&0&\ne0&\ne0
 \end{array},\qquad s_{\rm src}=w_{(s)}=0.
 \label{eq:phase_branches}
\end{equation}
The ordered fields, anisotropy $B(r)$ and horizon data are determined
by the same five-dimensional equations at each stationary point.

\section{Charged orders and the conformal reference theory}\label{sec:charged}
The purpose of the conformal reference is to decide which order is
thermodynamically favored when scalar and vector condensates first
become marginal at the same chemical potential. We begin with their
charged equations on the fixed ultraviolet background,
where the background is fixed and the condensate normalization
and thermodynamic ensemble can be specified explicitly. The
calculation proceeds from the source-free linear modes to the
fourth-order free energy, with the boundary conditions held fixed
within each ensemble. This one-parameter reference fixes the
quartic interactions exactly, but a transverse two-parameter
crossing belongs to the interpolating flow studied in
Section~\ref{sec:phase}.

\subsection{\texorpdfstring{Probe action and ordered ansatz}{Probe action and ordered ansatz}}
For the conformal reference we use the conventions
\eqref{eq:gauge_conventions} and the doublet orientation of the ordered
ansatz of Section~\ref{sec:reduction}. The two doublet orientations
are related by a constant $SU(2)$ rotation. The real amplitude
$\Phi$ of this reference is the ultraviolet limit of $s$, and $\omega$
is the ultraviolet limit of $w$. In the probe limit the charged part of
Eq.~\eqref{eq:S5}, written with the couplings
\eqref{eq:tetracritical_Z}, is
\begin{align}
 S_{\rm probe}&=\frac1{g_{\rm YM}^2}
 \int\dd^5x\sqrt{-g}\,L_{\rm probe},\qquad
 \frac1{g_{\rm YM}^2}=\frac{\eps}{2\kappa_5^2},
 \label{eq:probe_normalization}\\
 L_{\rm probe}&=-\frac14 Z(\phi)G^a_{\mu\nu}G^{a\mu\nu}
 -(D_{\mu}\boldsymbol\Phi)^\dagger
 D^{\mu}\boldsymbol\Phi
 -m_\Phi^2(\phi,f)\,\boldsymbol\Phi^\dagger\boldsymbol\Phi
 -\lambda_\Phi(\phi)(\boldsymbol\Phi^\dagger\boldsymbol\Phi)^2.
 \label{eq:gauge_action}
\end{align}
All conformal analytic formulas that follow use $Z=1$,
$m_\Phi^2=-4$, $\lambda_\Phi=\lambda_8$ and omit gravitational
response at leading order. The fully backreacted problem is governed
separately by the reduced action \eqref{eq:S5}.

For homogeneous scalar and vector orders in this UV reference, we take
\begin{equation}
 B=b(u)\tau^3\dd t+\omega(u)\tau^1\dd x_1,
 \qquad \boldsymbol\Phi=\frac1{\sqrt2}\begin{pmatrix}0\\\Phi(u)\end{pmatrix}.
 \label{eq:ansatz}
\end{equation}
In the exact UV probe reference, the nonzero covariant derivatives and field strengths are
\begin{align}
 D_u\boldsymbol\Phi&=\frac1{\sqrt2}\begin{pmatrix}0\\\Phi'\end{pmatrix},&
 D_t\boldsymbol\Phi&=\frac1{2\sqrt2}\begin{pmatrix}0\\-ib\Phi\end{pmatrix},&
 D_{x_1}\boldsymbol\Phi&=\frac1{2\sqrt2}\begin{pmatrix}i\omega\Phi\\0\end{pmatrix},
 \label{eq:explicit_covariant_derivatives}\\
 G^3_{ut}&=b',&G^1_{ux_1}&=\omega',&G^2_{tx_1}&=-b\omega.
 \label{eq:explicit_strengths}
\end{align}
Consequently
\begin{equation}
 (D_{\mu}\boldsymbol\Phi)^\dagger D^{\mu}\boldsymbol\Phi
 =\frac12(\partial\Phi)^2+\frac18\Phi^2 B_{\mu}^aB^{a\mu}.
 \label{eq:Higgs_mass_identity}
\end{equation}
The cross terms between $\partial_\mu\Phi$ and $B_\mu$ cancel
because $\Phi$ is real. The factor $1/8$ follows from $\{\tau^a,\tau^b\}=\delta^{ab}{\bf1}/2$ and $\boldsymbol\Phi^\dagger\boldsymbol\Phi=\Phi^2/2$. Thus the radial Lagrangian is exactly
\begin{equation}
 L_{\rm probe}\big|_{\rm ansatz}
 =-\frac14Z(\phi)G^a_{\mu\nu}G^{a\mu\nu}
 -\frac12(\partial\Phi)^2-\frac12m_\Phi^2\Phi^2
 -\frac18\Phi^2B_{\mu}^aB^{a\mu}
 -\frac{\lambda_\Phi}{4}\Phi^4.
 \label{eq:probe_action}
\end{equation}
The coefficient of the scalar--vector term is fixed by the fundamental representation and field normalization. The coefficient identities below retain $\lambda_\Phi(\phi)$ symbolically. In the ultraviolet AdS limit $Z(0)=1$, $m_\Phi^2(0)=-4$ and $\lambda_\Phi(0)=\lambda_8$, so the known exact BF-bound profiles remain valid reference solutions.

The ansatz consistently restricts the full field equations. The only nonzero Higgs gauge currents are $j^{3t}=\Phi^2B^{3t}/4$ and $j^{1x_1}=\Phi^2B^{1x_1}/4$, the upper doublet component is not sourced, and the radial gauge constraint vanishes. The full doublet action defines the theory even at the normal and pure-vector branches.
In particular,
\begin{equation}
 B_{\mu}^aB^{a\mu}=-\frac{b^2}{u^2 f_u}
 +\frac{\omega^2}{u^2}.
 \label{eq:B_squared}
\end{equation}
The temporal covariant derivative lowers the scalar quadratic potential, while the spatial derivative gives the positive interaction between the two condensates. Both effects come from the same gauge-invariant kinetic term.

\subsection{Radial field equations and source-free boundary data}\label{sec:radial_data}
In the ultraviolet $\AdS_5$ limit, $\phi=0$, the neutral winding
amplitude is $f=f_{\rm UV}$, and the conformal blackening function
is $f_u=1-u^{-4}$. The index on $f_u$ distinguishes the fixed
background function from the neutral winding amplitude.
Reduction of Eq.~\eqref{eq:probe_action} then gives the explicitly one-dimensional Lagrangian
\begin{align}
 L^{(1d)}={}&\frac12u^3b'^2-\frac12u^3f_u\omega'^2
 +\frac{b^2\omega^2}{2uf_u}
 -\frac12u^5f_u\Phi'^2-\frac12m_\Phi^2u^3\Phi^2\nonumber\\
 &+\frac{ub^2\Phi^2}{8f_u}-\frac{u\omega^2\Phi^2}{8}
 -\frac{\lambda_\Phi}{4}u^3\Phi^4.
 \label{eq:radial_action}
\end{align}
Its three independent Euler--Lagrange equations are
\begin{align}
 (u^3b')'&=\frac{b\omega^2}{uf_u}+\frac{ub\Phi^2}{4f_u},\label{eq:b_full}\\
 (u^3f_u\omega')'&=-\frac{b^2\omega}{uf_u}+\frac{u\Phi^2\omega}{4},\label{eq:w_full}\\
 (u^5f_u\Phi')'&=m_\Phi^2u^3\Phi-\frac{ub^2\Phi}{4f_u}
 +\frac{u\omega^2\Phi}{4}+\lambda_\Phi u^3\Phi^3.\label{eq:Phi_full}
\end{align}
Near $u=1$ one has $f_u=4(u-1)+O((u-1)^2)$ and $b=O(u-1)$.
Evaluating Eqs.~\eqref{eq:b_full}--\eqref{eq:Phi_full} at the horizon,
where $(u^3f_u\omega')'\to4\omega'(1)$ and
$(u^5f_u\Phi')'\to4\Phi'(1)$, gives the regularity conditions
\begin{align}
 \omega'(1)&=\frac{\Phi(1)^2\omega(1)}{16},&
 \Phi'(1)&=\frac{m_\Phi^2+\omega(1)^2/4+\lambda_\Phi\Phi(1)^2}{4}\Phi(1),\nonumber\\
 b''(1)&=b'(1)\left[\frac{\omega(1)^2+\Phi(1)^2/4}{4}-3\right].
 \label{eq:horizon_data}
\end{align}
At the AdS boundary, the source-free conditions at the BF mass are
\begin{align}
 b(\infty)&=b_\infty,& b(1)&=0,\label{eq:mu_normalization}\\
 \omega(u)&=\omega_{(s)}+w_0 u^{-2}+\cdots,& \omega_{(s)}&=0,\nonumber\\
 \Phi(u)&=u^{-2}[J_\Phi\log u+s_0+\cdots],&J_\Phi&=0.
 \label{eq:UV_sources}
\end{align}
Here $b_\infty$ is the Cartan chemical potential in our
normalization.\footnote{It is $\sqrt3$ times the chemical
potential of Ref.~\cite{Kim:2025hsv}.} The quartic derivation in Section~\ref{sec:perturbation} includes the finite scalar counterterm at the BF bound. The two consistent pure branches are $\Phi=0,\omega\neq0$ and $\omega=0,\Phi\neq0$, together with their common normal state. The boundary expansion at $m_\Phi^2=-4$ contains two degenerate powers, so the condition $J_\Phi=0$ on the logarithmic coefficient is imposed in addition to the leading falloff. This is the boundary condition used for the exact scalar zero mode \cite{BreitenlohnerFreedman:1982,KlebanovWitten:1999}. The two branches have different spatial symmetries, but they share the same action and the same boundary sources.

Let $\boldsymbol\Phi_{(s)}$ be the boundary doublet source and $\langle\boldsymbol O_\Phi\rangle$ its response. To derive the Ward identity, let $W_{\rm ren}[B_{(s)},\boldsymbol\Phi_{(s)}]$ be the renormalized generating functional. A boundary infinitesimal gauge transformation with parameter $\alpha=\alpha^a\tau^a$ acts on the sources as $\delta_\alpha B_{(s)i}=D_i\alpha$ and $\delta_\alpha\boldsymbol\Phi_{(s)}=-i\alpha\boldsymbol\Phi_{(s)}$. Using the definitions of the renormalized one-point functions and integrating the current term by parts gives
\begin{align}
 0=\delta_\alpha W_{\rm ren}
 =\int\!\dd^4x\sqrt{-\gamma_{(0)}}\,\alpha^a\Bigl[
 &-(D_i\langle J^i\rangle)^a
 +i\boldsymbol\Phi_{(s)}^\dagger\tau^a
     \langle\boldsymbol O_\Phi\rangle\nonumber\\[-2pt]
 &-i\langle\boldsymbol O_\Phi\rangle^\dagger
     \tau^a\boldsymbol\Phi_{(s)}\Bigr].
 \label{eq:Ward_variation}
\end{align}
Since $\alpha^a(x)$ is arbitrary, its coefficient must vanish.
\begin{equation}
 (D_i\langle J^i\rangle)^a
 =i\left[\boldsymbol\Phi_{(s)}^\dagger\tau^a\langle\boldsymbol O_\Phi\rangle
 -\langle\boldsymbol O_\Phi\rangle^\dagger\tau^a\boldsymbol\Phi_{(s)}\right].
 \label{eq:Ward_identity}
\end{equation}
At zero scalar source this is covariant current conservation. Since the only gauge source is a constant $B_t^3$, its third component obeys $\partial_i\langle J^{3i}\rangle=0$. The vector one-point function comes from the renormalized on-shell momentum conjugate to $B_{x_1}^1$. In radial gauge the Yang--Mills contribution at the AdS boundary is
\begin{equation}
 \Pi^{1x_1}=
 -\frac1{g_{\rm YM}^2}\sqrt{-g}\,Z(\phi)
 G^{1ux_1}=-\frac1{g_{\rm YM}^2}u^3f_u\omega'
 \qquad (\phi=0).
 \label{eq:current_momentum}
\end{equation}
Since $\omega=w_0u^{-2}+\cdots$ at vanishing source, its radial
derivative is $\omega'=-2w_0u^{-3}+\cdots$. Taking $u\to\infty$ yields
\begin{equation}
 \boxed{\;\langle J^{1x_1}\rangle
 =-\frac1{g_{\rm YM}^2}\lim_{u\to\infty}
 u^3f_u\omega'(u)=\frac{2}{g_{\rm YM}^2}w_0.\;}
 \label{eq:current_dictionary}
\end{equation}
This fixes the superfluid interpretation. The spatial vector order carries charge under the residual boundary $U(1)$ generated by $\tau^3$. The doublet component carries charge of magnitude $1/2$ under the same symmetry. A nonzero source-free condensate breaks the residual global symmetry in either thermodynamic ensemble \cite{Herzog:2009xv,Cai:2015cya,Bianchi:2001kw}.

\subsection{Fixed source and fixed Cartan charge}\label{sec:ensemble}
The electric boundary condition distinguishes the two thermodynamic
ensembles. The grand-canonical problem fixes the temporal source
$b_\infty$ in Eq.~\eqref{eq:mu_normalization}. Let
$S_{E,{\rm probe}}^{\rm ren}(T,b_\infty)$ be the renormalized
Euclidean on-shell action of the probe matter sector with this
Dirichlet boundary condition and with vanishing scalar and spatial
vector sources. If $V_3$ is the boundary spatial volume, the
grand-potential density is
$\Omega(T,b_\infty)=T S_{E,{\rm probe}}^{\rm ren}/V_3$.
The same neutral gravitational background is used to compare the
phases, and its common thermodynamic contribution cancels in their
free-energy differences. Multiplication by $g_{\rm YM}^2$ removes
the overall probe normalization from the radial formulas.

To pass to fixed Cartan density we vary the Euclidean on-shell
action with respect to $b_\infty$. The radial canonical momentum
of the temporal gauge field follows from
Eq.~\eqref{eq:radial_action},
\begin{equation}
 \pi_b(u)=\frac{\partial L^{(1d)}}{\partial b'(u)}
 =u^3b'(u),\qquad
 n=\lim_{u\to\infty}\pi_b(u).
 \label{eq:radial_flux_definition}
\end{equation}
On the backreacted flow the corresponding quantity is
$\pi_b=e^{2A+B}b'/X$ in the radial coordinate of
Eq.~\eqref{eq:backreacted_ansatz}, and $\rho$ in Eq.~\eqref{eq:thermo}
is $\eps$ times its boundary value. Charged condensates exchange flux
with the bulk, so $\pi_b(u)$ varies through an ordered interior, as
Eq.~\eqref{eq:b_full} shows. The derivation of the boundary
variation is explicit in the one-dimensional formulation. At fixed
temperature and fixed scalar and vector sources, the on-shell
variation of the radial charged action reduces to
\begin{equation}
 \delta\!\int_1^\infty\!\dd u\,L^{(1d)}
 =\int_1^\infty\!\dd u\,
 \left[\frac{\partial L^{(1d)}}{\partial b}
 -\frac{\dd}{\dd u}\left(\frac{\partial L^{(1d)}}{\partial b'}\right)\right]
 \delta b
 +\left[\pi_b\,\delta b\right]_1^\infty
 =n\,\delta b_\infty .
 \label{eq:electric_boundary_variation}
\end{equation}
The bulk term vanishes by Eq.~\eqref{eq:b_full}. The other field
variations have vanishing boundary sources. Regularity fixes
$b(1)=0$, hence $\delta b(1)=0$, while
$\delta b(\infty)=\delta b_\infty$. Under Euclidean continuation
$B_\tau^3=-ib$, the electric part of the on-shell Euclidean
functional contributes the opposite sign to the variation with
respect to real $b_\infty$. With $\beta=1/T$ and
$\Omega=T S_{E,{\rm probe}}^{\rm ren}/V_3$, this gives
\begin{equation}
 \delta S_{E,{\rm probe}}^{\rm ren}
 =-\frac{\beta V_3}{g_{\rm YM}^2}\,n\,\delta b_\infty,
 \qquad
 \delta\big(g_{\rm YM}^2\Omega\big)
 =\frac{g_{\rm YM}^2}{\beta V_3}
 \delta S_{E,{\rm probe}}^{\rm ren}.
 \label{eq:euclidean_variation_steps}
\end{equation}
The resulting grand-potential variation is
\begin{equation}
 \delta(g_{\rm YM}^2\Omega)
 =-n\,\delta b_\infty
 \quad\text{at fixed }T,\quad\boldsymbol\Phi_{(s)}=0,\quad
 \omega_{(s)}=0.
 \label{eq:grand_variation}
\end{equation}
The Legendre transform converts the Dirichlet variational problem for
$b_\infty$ into a Neumann problem for the asymptotic electric flux,
\begin{equation}
 n=-\partial_{b_\infty}(g_{\rm YM}^2\Omega),\qquad
 F(T,n)=g_{\rm YM}^2\Omega(T,b_\infty)+b_\infty n.
 \label{eq:canonical_ensemble}
\end{equation}
Here $F$ and $n$ are normalized probe quantities. The physical
Cartan density is $n_3=n/g_{\rm YM}^2$, and the physical Helmholtz
density is $F/g_{\rm YM}^2$. Differentiating
Eq.~\eqref{eq:canonical_ensemble} gives $\dd F|_T=b_\infty\dd n$,
because the two terms proportional to $\dd b_\infty$ cancel by
Eq.~\eqref{eq:grand_variation}.

The normal solution makes the normalization and the sign of the
transform explicit. With the horizon condition $b_N(1)=0$, the source
$b_\infty$ fixes $b_N=b_\infty(1-u^{-2})$ and hence
$u^3b_N'=2b_\infty$. The Maxwell contribution to the renormalized
potential is
\begin{equation}
 g_{\rm YM}^2\Omega_N
 =-\frac12\int_1^\infty\dd u\,u^3(b_N')^2
 =-\frac12 b_\infty n_N=-b_\infty^2.
 \label{eq:normal_on_shell_integral}
\end{equation}
Thus the normal-state data are
\begin{equation}
 g_{\rm YM}^2\Omega_N=-b_\infty^2,\qquad
 n_N=2b_\infty,\qquad\partial_{b_\infty}n_N=2.
 \label{eq:normal_density}
\end{equation}
In the canonical ensemble the same normal state becomes
\begin{equation}
 F_N(n)=\frac{n^2}{4},\qquad
 \frac{\partial F_N}{\partial n}=\frac n2=b_\infty,
 \qquad \frac{\partial^2F_N}{\partial n^2}=\frac12>0.
 \label{eq:normal_canonical_reference}
\end{equation}
The inverse of the last derivative is the normal charge
susceptibility $\chi_{\rm UV}=\partial_{b_\infty}n_N$, which controls
the canonical quartic coefficients in Section~\ref{sec:perturbation}.

The scalar and off-Cartan spatial sources vanish in both ensembles. We fix the conserved Cartan density $n$, without imposing local charge neutrality or introducing additional charged matter. The normal-state relation between $n_N$ and $b_\infty$ is monotone, so $n_N$ and temperature provide independent local thermodynamic controls. The Legendre transform must be performed before comparing the pure and mixed phases at the same density.

\subsection{Scalar--vector free energy in the conformal reference}\label{sec:perturbation}
\subsubsection{Zero modes and positive quadratic forms}\label{sec:critical_modes}
The ultraviolet limit of the eight-dimensional parent recovers the AdS$_5$
black brane at leading order. We use its probe sector as an exact
reference and solve this limit with
$m_\Phi^2(0)=-4$, $Z(0)=1$ and $\lambda_\Phi(0)=\lambda_8$.
The charge assignment and source-free boundary conditions agree
with those on the flow. In horizon units, $f_u=1-u^{-4}$ and the normal solution is
\begin{equation}
 b_N(u)=b_\infty(1-u^{-2}),\qquad \omega=\Phi=0.
 \label{eq:normal_solution}
\end{equation}
At the critical potential $b_\infty=b_c$, with $b_c=4$ in horizon
units, both source-free instabilities have exact radial modes.
\begin{align}
 (u^3f_u\psi')'+\frac{b_c^2(1-u^{-2})^2}{uf_u}\psi&=0,
 \label{eq:vector_zero_equation}\\
 (u^5f_u\Phi_1')'+4u^3\Phi_1+
 \frac{b_c^2u(1-u^{-2})^2}{4f_u}\Phi_1&=0,
 \label{eq:scalar_zero_equation}\\
 \psi(u)=\frac{u^2}{(1+u^2)^2},\qquad
 \Phi_1(u)&=\frac1{1+u^2}.
 \label{eq:zero_modes}
\end{align}
Their normalizable $u^{-2}$ coefficients are unity. The vector mode agrees
with Refs.~\cite{HerzogPufu,Basu:2011tt,Kim:2025hsv}. The scalar mode agrees
with Ref.~\cite{Herzog2010} after identifying its unit-charge potential with
$b/2$. The factor two follows from the doublet charge of magnitude $1/2$.
Both profiles are nodeless and regular, with $\psi'(1)=0$ and
$\Phi_1'(1)=-\Phi_1(1)$.

Their status as the lowest modes follows directly from positive quadratic
forms. For regular source-free $y$,
\begin{equation}
 \int_1^\infty\!\dd u\left[u^3f_uy'^2-
 \frac{b_c^2(1-u^{-2})^2}{uf_u}y^2\right]
 =\int_1^\infty\!\dd u\,u^3f_u\psi^2
 \left[\left(\frac y\psi\right)'\right]^2\geq0.
 \label{eq:vector_positive_form}
\end{equation}
The corresponding scalar identity includes the finite BF boundary term.
\begin{align}
 &\lim_{R\to\infty}\left\{
 \int_1^R\!\dd u\left[u^5f_uy'^2-4u^3y^2
 -\frac{b_c^2u(1-u^{-2})^2}{4f_u}y^2\right]
 +2R^4y(R)^2\right\}\nonumber\\[-1mm]
 &\hspace{12mm}=\int_1^\infty\!\dd u\,u^5f_u\Phi_1^2
 \left[\left(\frac y{\Phi_1}\right)'\right]^2\geq0.
 \label{eq:scalar_positive_form}
\end{align}
These identities concern the homogeneous radial sector with
$J_\Phi=0$ and no double-trace deformation. In this exact limit only
the state parameter $\delta b=b_\infty-b_c$ is varied, at fixed scalar
mass and boundary condition.

\subsubsection{Electric response, renormalized quartics and fixed flux}\label{sec:electric_feedback}
Assign $\delta b$ degree two in the condensate amplitudes and write
\begin{align}
 \omega&=w_0\psi+\omega_3+O(5),\qquad
 \Phi=s_0\Phi_1+\Phi_3+O(5),\nonumber\\
 b&=b_\infty(1-u^{-2})+
 w_0^2b_2^\omega+s_0^2b_2^\Phi+O(4).
 \label{eq:field_expansion}
\end{align}
The cubic profiles are constructed in Appendix~\ref{app:cubic}.
Inserting Eq.~\eqref{eq:field_expansion} into Eq.~\eqref{eq:b_full}
gives $(u^3b_2^{i\prime})'=S_i$ with the electric sources
\begin{equation}
 S_\omega=\frac{b_c(1-u^{-2})\psi^2}{uf_u},\qquad
 S_\Phi=\frac{b_c\,u(1-u^{-2})\Phi_1^2}{4f_u}.
 \label{eq:electric_sources}
\end{equation}
With the zero modes \eqref{eq:zero_modes} the induced electric
fields are
\begin{align}
 (u^3 b_2^{\omega\prime})'&=\frac{4u^5}{(1+u^2)^5},&
 b_2^\omega(u)&=-\frac{(u^2-1)(3u^4+12u^2+5)}
 {96u^2(1+u^2)^3},
 \label{eq:b2I_solution}\\
 (u^3 b_2^{\Phi\prime})'&=\frac{u^3}{(1+u^2)^3},&
 b_2^\Phi(u)&=-\frac{u^2-1}{16u^2(1+u^2)}.
 \label{eq:b2II_solution}
\end{align}
Both vanish at the horizon and boundary and are negative in the interior.
Condensation therefore screens the electric potential at fixed
$b_\infty$.

The quartic interaction is obtained from the renormalized Euclidean
action. Let $S_E^{\rm ren}$ denote that action and $V_3$ the boundary
spatial volume. Set $\Omega=TS_E^{\rm ren}/V_3$ and define
\begin{equation}
 \Delta\Omega\equiv g_{\rm YM}^2(\Omega-\Omega_N).
 \label{eq:Omega_definition}
\end{equation}
With $B^3_\tau=-ib$, the Euclidean functional in the real potential $b$ is
\begin{align}
 g_{\rm YM}^2\Omega_{\rm probe}
 =\lim_{R\to\infty}\bigg\{\int_1^R\!\dd u\bigg[&
 -\frac12u^3b'^2+\frac12u^3f_u\omega'^2-\frac{b^2\omega^2}{2uf_u}
 +\frac12u^5f_u\Phi'^2-2u^3\Phi^2\nonumber\\
 &-\frac{ub^2\Phi^2}{8f_u}+\frac18u\omega^2\Phi^2
 +\frac{\lambda_\Phi}{4}u^3\Phi^4\bigg]
 +g_{\rm YM}^2\Omega_{\rm ct}(R)\bigg\},
 \label{eq:renormalized_functional}\\
 g_{\rm YM}^2\Omega_{\rm ct}(R)
 &=R^4\sqrt{f_u(R)}\Phi(R)^2
 \left(1-\frac1{2\log R}\right).
 \label{eq:counterterms}
\end{align}
The BF Dirichlet counterterm is gauge invariant because
$\Phi^2=2\boldsymbol\Phi^\dagger\boldsymbol\Phi$
\cite{Bianchi:2001kw,Skenderis:2002wp,Herzog2010}. Derivative counterterms
vanish for the sources considered. At zero scalar source the inverse-logarithm
term vanishes in the limit, but the first term remains finite. In particular,
\begin{equation}
 \int_1^\infty\left[u^5f_u\Phi_1'^2-4u^3\Phi_1^2
 -\frac{b_c^2u(1-u^{-2})^2}{4f_u}\Phi_1^2\right]\dd u=-2.
 \label{eq:BF_cancellation}
\end{equation}
The BF boundary contribution is $2\lim_{R\to\infty}R^4\Phi_1(R)^2=2$.
Their cancellation is necessary for the scalar zero mode. The normal
potential is $g_{\rm YM}^2\Omega_{N,\rm probe}=-b_\infty^2$.

The quadratic coefficients follow from the $\delta b$ dependence of
Eq.~\eqref{eq:renormalized_functional}. Since
$b_N^2=b_c^2(1-u^{-2})^2+2b_c\,\delta b\,(1-u^{-2})^2+O(\delta b^2)$,
the zero-mode equations remove the terms independent of $\delta b$, and
the remainder is $a_\omega=-Q_\omega\delta b$ and $a_\Phi=-Q_\Phi\delta b$
with
\begin{equation}
 Q_\omega=\int_1^\infty(1-u^{-2})\,S_\omega\,\dd u,
 \qquad
 Q_\Phi=\int_1^\infty(1-u^{-2})\,S_\Phi\,\dd u.
 \label{eq:projection_integrals}
\end{equation}
Let $G_{\rm UV}(u,u')=\frac12(1-u_<^{-2})\,u_>^{-2}$ be the Green
function of $-(u^3\partial_u)'$ with Dirichlet conditions at the horizon
and the boundary, where $u_<$ and $u_>$ are the smaller and larger of
$u,u'$. Then $b_2^i=-\int G_{\rm UV}S_i$, and the induced flux is
$\lim_{u\to\infty}u^3b_2^{i\prime}=Q_i$. The coefficient $Q_i$ is
therefore both the slope of the quadratic coefficient and the induced
Cartan flux per unit squared amplitude, as required by
Eq.~\eqref{eq:grand_variation}.
Terms linear in the cubic profiles vanish by the zero-mode equations and
the renormalized source-free condition. Integrating the electric equation
by parts leaves
\begin{align}
 \Delta\Omega^{(4)}={}&
 \frac12\int_1^\infty\!\dd u\,u^3
 (w_0^2b_2^{\omega\prime}+s_0^2b_2^{\Phi\prime})^2
 +\frac{w_0^2s_0^2}{8}\int_1^\infty\!\dd u\,u\psi^2\Phi_1^2
 \nonumber\\
 &+\frac{\lambda_\Phi s_0^4}{4}
 \int_1^\infty\!\dd u\,u^3\Phi_1^4.
 \label{eq:quartic_integral}
\end{align}
The temporal equation fixes the positive electric contribution.
Indeed, multiplying each equation in
Eqs.~\eqref{eq:b2I_solution}--\eqref{eq:b2II_solution}
by its own Dirichlet correction and integrating by parts gives
\begin{equation}
 \int_1^\infty\!\dd u\,b_2^i(u)(u^3(b_2^i)'(u))'
 =-\int_1^\infty\!\dd u\,u^3[(b_2^i)'(u)]^2,
 \quad i\in\{\omega,\Phi\}.
 \label{eq:analytic_electric_energy_identity}
\end{equation}
The Dirichlet boundary condition gives Eq.~\eqref{eq:quartic_integral}.

Define the electric overlap integrals and the two contact integrals
\begin{equation}
 \begin{gathered}
 I_{ij}=\int_1^\infty u^3\,b_2^{i\prime}b_2^{j\prime}\,\dd u
 =\int_1^\infty\!\!\int_1^\infty S_i(u)G_{\rm UV}(u,u')S_j(u')\,\dd u\,\dd u',\\
 H=\frac18\int_1^\infty u\psi^2\Phi_1^2\,\dd u,\qquad
 A_\Phi=\frac14\int_1^\infty u^3\Phi_1^4\,\dd u .
 \end{gathered}
 \label{eq:overlap_integrals}
\end{equation}
The quartic coefficients are defined directly by the radial response,
\begin{equation}
 b_\omega^{(\mu)}=\frac12I_{\omega\omega},\qquad
 b_\Phi^{(\mu)}=\frac12I_{\Phi\Phi}+\lambda_\Phi A_\Phi,\qquad
 g^{(\mu)}=I_{\omega\Phi}+H.
 \label{eq:analytic_all_quartic_terms}
\end{equation}
The superscript $(\mu)$ denotes fixed boundary Cartan potential.
The first two terms measure how each condensate changes the electric
field, and the remaining terms represent direct scalar interactions.
The electric overlap matrix $I_{ij}$ is nonnegative, while $H$
and $A_\Phi$ are positive. Thus Eq.~\eqref{eq:analytic_all_quartic_terms}
connects the sign of the mixed-phase interaction to radial profiles
rather than to independently prescribed phenomenological constants.
Appendix~\ref{app:cubic} gives the exact analytic derivation without
introducing explicit coefficient tables.

Combining them yields
\begin{equation}
 \Delta\Omega=-\delta b\left(Q_\omega w_0^2+Q_\Phi s_0^2\right)
 +b_\omega^{(\mu)}w_0^4+b_\Phi^{(\mu)}s_0^4
 +g^{(\mu)}w_0^2s_0^2+O(6).
 \label{eq:complete_potential}
\end{equation}
Its stationarity conditions agree with the independent cubic projections.
\begin{align}
 0&=w_0\left[-Q_\omega\delta b+2b_\omega^{(\mu)}w_0^2
 +g^{(\mu)}s_0^2\right],\label{eq:vector_amplitude}\\
 0&=s_0\left[-Q_\Phi\delta b+2b_\Phi^{(\mu)}s_0^2
 +g^{(\mu)}w_0^2\right].\label{eq:scalar_amplitude}
\end{align}
The scalar electric contribution agrees with the minimal model of Ref.~\cite{Natsuume:2025GL} under
$u_{\rm N}=u^{-2}$, $A_t^{\rm N}=b/2$ and
$\Psi^{\rm N}=-\Phi/(2\sqrt2)$, with our scalar action four times its
unit-charge normalization. The mixed coefficient also requires the direct
covariant-derivative interaction.

The same radial calculation can be performed at fixed Cartan density.
Differentiating Eq.~\eqref{eq:complete_potential} with
Eq.~\eqref{eq:grand_variation} gives the Cartan density
\begin{equation}
 n=\chi_{\rm UV}b_\infty+Q_\omega w_0^2+Q_\Phi s_0^2+\cdots,
 \qquad \chi_{\rm UV}=\partial_{b_\infty}n_N=2.
 \label{eq:induced_charge}
\end{equation}
At fixed $n$ the potential shifts to
$b_\infty=n/\chi_{\rm UV}-(Q_\omega w_0^2+Q_\Phi s_0^2)/\chi_{\rm UV}$.
Inserting this shift into $F=g_{\rm YM}^2\Omega+b_\infty n$ with
$g_{\rm YM}^2\Omega_N=-\chi_{\rm UV}b_\infty^2/2$, the terms linear in
the shift cancel and the normal state at the same density is
subtracted. The Legendre transform therefore adds
\begin{equation}
 \Delta F^{(4)}=\Delta\Omega^{(4)}+
 \frac{1}{2\chi_{\rm UV}}\left(Q_\omega w_0^2+Q_\Phi s_0^2\right)^2.
 \label{eq:rankone_electric}
\end{equation}
Its quadratic coefficients are $a_i=-Q_i(n-n_c)/\chi_{\rm UV}$ with
$n_c=\chi_{\rm UV}b_c$, and its quartics are
\begin{equation}
 b_\omega^{(n)}=b_\omega^{(\mu)}+\frac{Q_\omega^2}{2\chi_{\rm UV}},\qquad
 b_\Phi^{(n)}=b_\Phi^{(\mu)}+\frac{Q_\Phi^2}{2\chi_{\rm UV}},\qquad
 g^{(n)}=g^{(\mu)}+\frac{Q_\omega Q_\Phi}{\chi_{\rm UV}}.
 \label{eq:canonical_coefficients}
\end{equation}
The fixed-flux electric solutions provide a radial check.
\begin{equation}
 b_{2,n}^i=b_2^i-\frac{Q_i}{\chi_{\rm UV}}(1-u^{-2}),\qquad
 i\in\{\omega,\Phi\}.
 \label{eq:fixed_flux_profiles}
\end{equation}
Their boundary fluxes vanish. The exact integrals in
Appendix~\ref{app:cubic} reproduce Eq.~\eqref{eq:canonical_coefficients},
including the same Higgs contact and scalar self-interaction terms.
This provides a radial check of the Legendre transform \cite{Yin:2013}.

For either ensemble, the scalar and vector condensates are governed
by the quartic potential
\begin{equation}
 \Delta F=a_\omega w_0^2+a_\Phi s_0^2+
 b_\omega w_0^4+b_\Phi s_0^4+
 gw_0^2s_0^2.
 \label{eq:abstract_landau}
\end{equation}
Its quartic matrix is the symmetric matrix with diagonal entries
$2b_\omega,2b_\Phi$ and off-diagonal entry $g$. We denote it
by $\mathbf B_\mu$ at fixed chemical potential in
Eq.~\eqref{eq:quartic}.
For a mixed stationary solution,
\begin{equation}
 w_0^2=\frac{g a_\Phi-2b_\Phi a_\omega}{4b_\omega b_\Phi-g^2},
 \qquad
 s_0^2=\frac{g a_\omega-2b_\omega a_\Phi}{4b_\omega b_\Phi-g^2},
 \label{eq:explicit_stationarity}
\end{equation}
and the Hessian in the ordinary amplitudes satisfies
\begin{equation}
 \det\big[\partial^2_{(w_0,s_0)}\Delta F\big]=16w_0^2s_0^2(4b_\omega b_\Phi-g^2).
 \label{eq:general_mixed_hessian}
\end{equation}
A stable mixed state requires $b_\omega,b_\Phi>0$,
both squared amplitudes to be positive, and the quartic
determinant to satisfy
\begin{equation}
 4b_\omega b_\Phi-g^2>0.
 \label{eq:quartic_stability}
\end{equation}
These requirements follow from the amplitude Hessian
\eqref{eq:general_mixed_hessian} and the explicit stationary
solution \eqref{eq:explicit_stationarity}, using coefficients
from the same chosen ensemble. In the conformal reference
both quadratic coefficients depend on the same parameter
$\delta b$, through
$(a_\omega,a_\Phi)=-(Q_\omega,Q_\Phi)\delta b$.
This single direction does not describe a two-dimensional
coexistence region. On a general background one needs two
independent thermodynamic controls to move through the distinct
normal, pure and mixed phases. Section~\ref{sec:phase} makes those
conditions explicit without assuming that the full-flow double
onset has already been established.

\subsubsection{\texorpdfstring{Ultraviolet thresholds and the canonical shift}{Ultraviolet thresholds and the canonical shift}}
The scalar quartic coupling enters only $b_\Phi$, through the positive
coefficient $A_\Phi$ in Eq.~\eqref{eq:analytic_all_quartic_terms}.
Each determinant is therefore affine in $\lambda_8$,
\begin{equation}
 \det\mathbf B_{\rm UV}^{(\sigma)}=4b_\omega^{(\sigma)}A_\Phi
 \big(\lambda_8-\lambda_c^{(\sigma)}\big),\qquad
 \lambda_c^{(\sigma)}=\frac{[g^{(\sigma)}]^2
 -4b_\omega^{(\sigma)}\big(b_\Phi^{(\sigma)}-\lambda_8A_\Phi\big)}
 {4b_\omega^{(\sigma)}A_\Phi},\qquad \sigma\in\{\mu,n\},
 \label{eq:cartan_formal_threshold}
\end{equation}
and $\lambda_c^{(\sigma)}$ is independent of $\lambda_8$. Inserting the
canonical coefficients \eqref{eq:canonical_coefficients}, the terms
containing $b_\Phi^{(\mu)}$ and those of second order in $1/\chi$ cancel
in the numerator of $\lambda_c^{(n)}-\lambda_c^{(\mu)}$, and the
remainder is minus a perfect square,
\begin{equation}
 \boxed{\;\lambda_c^{(n)}=\lambda_c^{(\mu)}
 -\frac{\big(g^{(\mu)}Q_\omega-2b_\omega^{(\mu)}Q_\Phi\big)^2}
 {4A_\Phi\,b_\omega^{(\mu)}\big(2\chi\,b_\omega^{(\mu)}+Q_\omega^2\big)}
 \;\le\;\lambda_c^{(\mu)} .\;}
 \label{eq:threshold_shift}
\end{equation}
Here $\chi=\chi_{\rm UV}$. The derivation uses only the rank-one form of
Eq.~\eqref{eq:rankone_electric}, so the same identity holds at any double
onset where the electric flux relaxes in this way, with the local
susceptibility in place of $\chi_{\rm UV}$.

The canonical threshold therefore lies below the grand-canonical one.
For $\lambda_8<\lambda_c^{(n)}$ both ultraviolet determinants are
negative, and the conformal double onset has no locally stable mixed
minimum in either ensemble. For
$\lambda_c^{(n)}<\lambda_8<\lambda_c^{(\mu)}$ only the canonical quartic
matrix is positive. Determining stability in this interval also
requires the coupled density--order variation, which is treated in
Section~\ref{sec:phase}.

\section{Infrared modes and scalar--vector competition}\label{sec:phase}
The locally $\AdS_8$ region fixes the infrared scalar and vector
onset scales. Their coincidence selects one effective scalar mass.
Thermodynamic competition on the complete flow is determined by
the ultraviolet source conditions and the quartic interactions
of the two charged modes.

\subsection{Infrared spectrum and the double onset}
At zero temperature the locally $\AdS_8$ infrared region lies at
$r\to0$. If a low-temperature black-brane family approaches this
flow, its horizon at $r=r_h>0$ cuts off that endpoint. When the
horizon lies well inside the infrared region, the leading
near-horizon geometry is the eight-dimensional AdS--Schwarzschild
black brane. We study the electrically probed onset in this
scaling limit. With $v=r/r_h$ and $h(v)=1-v^{-7}$, the regular
Cartan potential is $b_0=\mu(1-v^{-5})$.
The dimensionless scalar mass squared and chemical potential are
\begin{equation}
 M=L_8^2m_0^2-56\xi
 =L_8^2(-4+20\xi-3\eta f_{\rm UV}^2)-56\xi,
 \qquad \widehat\mu=\frac{L_8^2\mu}{r_h}.
 \label{eq:IR_M}
\end{equation}
The horizon-regular, source-free scalar and vector perturbations
satisfy
\begin{align}
 -\big(v^6h\,w_1'\big)'&=
 \widehat\mu_p^2\frac{v^2(1-v^{-5})^2}{h}\,w_1,
 \label{eq:IR_SL_p}\\
 -\big(v^8h\,s_1'\big)'+Mv^6s_1&=
 \frac{\widehat\mu_s^2}{4}
 \frac{v^4(1-v^{-5})^2}{h}\,s_1.
 \label{eq:IR_SL_s}
\end{align}
The two eigenfunctions have boundary falloffs
$w_1\sim v^{-5}$ and $s_1\sim v^{-\Delta}$, where
\begin{equation}
 \Delta(\Delta-7)=M,\qquad \Delta>\frac72.
 \label{eq:IR_dimension}
\end{equation}
This is the $\AdS_8$ mass relation for a mode that is constant on
the torus. Equation~\eqref{eq:charged_IR_BF} is precisely the
condition that $\Delta$ be real. The factor $1/4$ in
\eqref{eq:IR_SL_s} comes from the charge of the fundamental
$SU(2)$ doublet.

The lowest eigenvalues have the variational representations
\begin{align}
 \widehat\mu_p^2&=\inf_{w\ne0}
 \frac{\int_1^\infty v^6h|w'|^2\,\dd v}
 {\int_1^\infty v^2(1-v^{-5})^2h^{-1}|w|^2\,\dd v},
 \nonumber\\
 \frac{\widehat\mu_s^2}{4}&=\inf_{s\ne0}
 \frac{\int_1^\infty\big(v^8h|s'|^2+Mv^6|s|^2\big)\dd v}
 {\int_1^\infty v^4(1-v^{-5})^2h^{-1}|s|^2\,\dd v}.
 \label{eq:IR_Rayleigh}
\end{align}
The fast falloff and regular horizon conditions are understood.
For a simple lowest scalar eigenvalue the Hellmann--Feynman identity
gives
\begin{equation}
 \frac{\partial\widehat\mu_s^2}{\partial M}
 =4\frac{\int_1^\infty v^6|s_1|^2\,\dd v}
 {\int_1^\infty v^4(1-v^{-5})^2h^{-1}|s_1|^2\,\dd v}>0.
 \label{eq:IR_monotonicity}
\end{equation}
The scalar onset moves through the vector onset as $M$ increases.
Inserting $s=v^{-\Delta}$ into Eq.~\eqref{eq:IR_Rayleigh} gives
\begin{equation}
 \widehat\mu_s^2\le
 \frac{2\Delta}{\displaystyle\int_0^1
 x^{2\Delta-6}\frac{(1-x^5)^2}{1-x^7}\,\dd x}
 <\frac{(2\Delta)^2[(2\Delta)^2-25]}{50},
 \qquad \Delta>\frac72 .
 \label{eq:IR_scalar_trial_bound}
\end{equation}
Here $x=1/v$, and the last bound uses $(1-x^7)^{-1}>1$.
For the vector profile $w_{\rm tr}=x^5/(1+x^2)^4$, direct
substitution into the radial operator gives
\begin{equation}
 -\partial_x\!\left[\frac{1-x^7}{x^4}\partial_xw_{\rm tr}\right]
 \ge 24\,\frac{(1-x^5)^2}{x^4(1-x^7)}w_{\rm tr},
 \qquad 0<x<1.
 \label{eq:IR_vector_comparison}
\end{equation}
Writing the vector quadratic form in terms of $w/w_{\rm tr}$
leaves a nonnegative derivative square and a vanishing boundary
term, giving $\widehat\mu_p^2\ge24$. The scalar variational
bound lies strictly below this vector bound as $M$ approaches
the $\AdS_8$ BF value from above.
At $M=-6$, the scalar has the fast falloff $s\sim v^{-6}$,
so $s=w/v$ maps admissible vector profiles to scalar profiles
with the same regularity and boundary conditions. This substitution
also makes the two denominators in \eqref{eq:IR_Rayleigh} identical.
Expanding the scalar derivative term and integrating its cross term
by parts gives the vector numerator plus
$\int_1^\infty v^{-3}|w|^2\,\dd v$. The boundary contribution
vanishes at the horizon and at infinity. The extra term is
strictly positive for the lowest scalar eigenfunction, so
minimizing the two quotients gives
$\widehat\mu_s^2(-6)>4\widehat\mu_p^2$. Hence the scalar
eigenvalue is larger at $M=-6$, whereas the bounds above place
it below the vector eigenvalue near the $\AdS_8$ BF limit.
The scalar eigenvalue varies continuously and increases strictly with $M$ by
\eqref{eq:IR_monotonicity}, while the vector eigenvalue does not
depend on $M$. They can therefore coincide at exactly one mass,
\begin{equation}
 -\frac{49}{4}<M_*<-6,\qquad
 \widehat\mu_s(M_*)=\widehat\mu_p .
 \label{eq:IR_mass_crossing}
\end{equation}

This mass fixes a relation among the parent couplings while
retaining the ultraviolet mass condition \eqref{eq:parameters}.
For an admissible neutral vacuum with $\lambda_\Psi\ge0$,
Eq.~\eqref{eq:vacuum_relations} implies $L_8^2\ge7/2$.
One possible fixed choice is
\begin{equation}
 \eta=0,\qquad
 \xi=\frac{M_*+4L_8^2}{20L_8^2-56},\qquad
 m_0^2=-4+20\xi .
 \label{eq:IR_parent_mass_matching}
\end{equation}
These constants give $m_{\rm UV}^2=-4$ and $M=M_*$.
The normal state has $\mathcal F=1$, so its positive gravitational
kinetic coefficient persists sufficiently close to onset.
The couplings are chosen once and held fixed as the state changes.
The infrared onsets are rays in the $(T,\mu)$ plane because
$T=7r_h/(4\pi L_8^2)$ and $\widehat\mu=L_8^2\mu/r_h$.
At $M=M_*$ these rays coincide. On the interpolating geometry,
the regular perturbations with vanishing ultraviolet sources define
$a_\omega(T,\mu)$ and $a_\Phi(T,\mu)$ at fixed neutral source
$J$. A common onset with distinct tangent directions satisfies
\begin{equation}
 a_\omega(T_\star,\mu_\star)=a_\Phi(T_\star,\mu_\star)=0,
 \qquad
 \det\begin{pmatrix}
 \partial_Ta_\omega&\partial_\mu a_\omega\\
 \partial_Ta_\Phi&\partial_\mu a_\Phi
 \end{pmatrix}_{(T_\star,\mu_\star)}\ne0 .
 \label{eq:critical}
\end{equation}
The two vanishing coefficients specify simultaneous normalizable
zero modes on one normal background. The nonzero Jacobian ensures
that their onset curves have different tangent directions in the
$(T,\mu)$ plane. Evaluating these coefficients requires the complete
neutral and charged radial profiles at the same $J$ and parent
couplings.

If a full-flow common onset is found, the nearby charged fields
can be expanded as $w=w_0w_\star+\cdots$ and
$s=s_0s_\star+\cdots$. The limits $s_0=0$ and $w_0=0$
select the pure $p$ and pure $s$ directions, respectively.
Nonzero values of both amplitudes describe the candidate
mixed continuation.

\subsection{Quartic form from nonlinear continuation}
The horizon amplitudes are convenient local order coordinates.
Their nonzero conversion to ultraviolet condensates does not
change the phase topology. Write
\begin{equation}
 \frac{\Omega-\Omega_N}{\eps}
 =a_\omega w_0^2+a_\Phi s_0^2+
 \frac12\begin{pmatrix}w_0^2&s_0^2\end{pmatrix}
 {\bf B}_\mu
 \begin{pmatrix}w_0^2\\s_0^2\end{pmatrix}
 +O\big((w_0^2+s_0^2)^3\big).
 \label{eq:Landau}
\end{equation}
The coefficients $a_i$ are the renormalized quadratic terms
of the normal state evaluated on the lowest modes. Near a
transverse crossing they vary as
\begin{align}
 a_\omega&=(\partial_Ta_\omega)_\star(T-T_\star)
 +(\partial_\mu a_\omega)_\star(\mu-\mu_\star)+\cdots,
 \nonumber\\
 a_\Phi&=(\partial_Ta_\Phi)_\star(T-T_\star)
 +(\partial_\mu a_\Phi)_\star(\mu-\mu_\star)+\cdots.
 \label{eq:slopes}
\end{align}
Their linear independence is the determinant condition in
Eq.~\eqref{eq:critical}. The renormalized fourth-order terms
are organized in a symmetric matrix with the entries of
Eq.~\eqref{eq:abstract_landau},
\begin{equation}
 {\bf B}_\mu=
 \begin{pmatrix}2b_\omega&g\\g&2b_\Phi\end{pmatrix},
 \qquad b_\omega>0,\quad b_\Phi>0,\quad
 \det{\bf B}_\mu=4b_\omega b_\Phi-g^2>0.
 \label{eq:quartic}
\end{equation}
At a common onset satisfying Eq.~\eqref{eq:critical}, the
inequalities in Eq.~\eqref{eq:quartic} determine mixed-order
stability.
Appendix~\ref{app:gravitational_quartics} specifies the coupled
Einstein--matter responses that determine its entries and
shows how the boundary ensemble enters the calculation.

If the conditions above hold, four local phase boundaries
follow from Eq.~\eqref{eq:Landau}. The two normal-state
boundaries are $a_\omega=0$ and $a_\Phi=0$. On a pure $p$
solution, $w_0^2=-a_\omega/(2b_\omega)$, and the scalar
instability satisfies
\begin{equation}
 a_\Phi-\frac{g}{2b_\omega}a_\omega=0 .
 \label{eq:secondary_p}
\end{equation}
On a pure $s$ solution the corresponding condition is
\begin{equation}
 a_\omega-\frac{g}{2b_\Phi}a_\Phi=0 .
 \label{eq:secondary_s}
\end{equation}
In the mixed region, $(w_0^2,s_0^2)^T=-{\bf B}_\mu^{-1}(a_\omega,a_\Phi)^T$ has positive entries. Positivity of \eqref{eq:quartic}, together with the nonzero Jacobian, gives the local tetracritical topology and continuous boundaries. These are local statements about the renormalized action under the displayed spectral hypotheses.

Figure~\ref{fig:phase} shows this local topology in the normal-state
quadratic coefficients. When Eq.~\eqref{eq:critical} holds, these
coordinates are locally equivalent to $(T,\mu)$.
\begin{figure}[htbp]
 \centering
 \begin{tikzpicture}[>=Latex,font=\small,scale=0.96]
  \fill[gray!14] (-2.4,-1.55) rectangle (0,0);
  \fill[blue!12] (0,0)--(5,0)--(5,1.7)--cycle;
  \fill[teal!25] (0,0)--(5,1.7)--(5,2.65)--cycle;
  \fill[orange!13] (0,0)--(5,2.65)--(5,4)--(0,4)--cycle;
  \fill[orange!13] (-2.4,0) rectangle (0,4);
  \draw[->,thick] (-2.55,0)--(5.25,0) node[right] {$-a_\omega$};
  \draw[->,thick] (0,-1.7)--(0,4.15) node[above] {$-a_\Phi$};
  \draw[very thick,blue!75!black] (0,0)--(5,1.7);
  \draw[very thick,orange!75!black] (0,0)--(5,2.65);
  \fill (0,0) circle (1.8pt);
  \node at (-1.35,-0.8) {normal};
  \node at (3.4,0.62) {$p$};
  \node at (-1.35,2.0) {$s$};
  \node at (3.9,2.05) {$s+p$};
  \node[below left] at (0,0) {$*$};
 \end{tikzpicture}
 \caption{Local phase topology for a positive quartic form with
 competing orders ($g>0$) and transverse scalar/vector onsets.
 The two rays are the secondary instabilities obtained from
 \eqref{eq:secondary_p}--\eqref{eq:secondary_s}. This schematic
 illustrates the analytic conditions and does not assert a
 realized crossing on the interpolating background.}
 \label{fig:phase}
\end{figure}

The free-energy ordering near a crossing, when it exists, follows
directly from the quartic functional. The pure branches obey
$\Delta\Omega_p/\eps=-a_\omega^2/(4b_\omega)$ and
$\Delta\Omega_s/\eps=-a_\Phi^2/(4b_\Phi)$. Substituting the mixed
solution into the same functional gives
\begin{align}
 \frac{\Omega_{s+p}-\Omega_p}{\eps}
 &=-\frac{\big(2b_\omega a_\Phi-g\,a_\omega\big)^2}
 {4b_\omega\det\mathbf B_\mu}+O(\Vert a\Vert^3),
 \nonumber\\
 \frac{\Omega_{s+p}-\Omega_s}{\eps}
 &=-\frac{\big(2b_\Phi a_\omega-g\,a_\Phi\big)^2}
 {4b_\Phi\det\mathbf B_\mu}+O(\Vert a\Vert^3).
 \label{eq:analytic_gaps}
\end{align}
Both denominators are positive precisely when the quartic form is
positive. In the interior of the coexistence wedge the differences
are strictly negative sufficiently close to the common onset.
The numerators vanish on the secondary boundaries
\eqref{eq:secondary_p} and \eqref{eq:secondary_s}, where the mixed
branch joins the pure branches continuously.
Section~\ref{sec:parent_checks} gives the nonlinear first-law
integral for extending this comparison to finite amplitudes.

The coefficients of Eq.~\eqref{eq:quartic} can be computed from the
radial equations when the corresponding normal background is known.
At the common onset, the charged fields and the Cartan potential
expand as
\begin{align}
 w&=w_0w_\star+O(w_0^3,w_0s_0^2),&
 s&=s_0s_\star+O(s_0^3,s_0w_0^2),\nonumber\\
 b&=b_N+w_0^2\delta b_\omega+s_0^2\delta b_s
 +O\big((w_0^2+s_0^2)^2\big).
 \label{eq:metric_response_expansion}
\end{align}
For the metric of Eq.~\eqref{eq:backreacted_ansatz}, the exact Cartan
equation in Eq.~\eqref{eq:matterODE} takes the flux form
\begin{equation}
 \left(\frac{e^{2A+B}}{X}b'\right)'=
 \frac{e^{-B}}{Xh}\,b w^2
 +\frac{e^{2A+B}}{4h}\,b s^2 .
 \label{eq:full_cartan_flux}
\end{equation}
The induced electric profiles follow by collecting the coefficients
of $w_0^2$ and $s_0^2$ in this equation. The metric functions,
breathing mode and neutral scalar also respond at this order.
their explicit contributions to the electric equations and the
fixed-source boundary conditions are displayed in
Appendix~\ref{app:gravitational_quartics}. At the conformal probe
endpoint the electric profiles are precisely
Eqs.~\eqref{eq:b2I_solution}--\eqref{eq:b2II_solution}, and their
radial integrals give Eq.~\eqref{eq:analytic_all_quartic_terms}.
On the fully backreacted flow, the neutral and metric responses
enter the quartic coefficients at order $\eps$.
Their contributions cannot in general be obtained by copying
the endpoint overlaps. The dependence on the parent quartic
coupling is derived in Appendix~\ref{app:phase_criterion}.

\subsection{Fixed-density free energy and ultraviolet--infrared stability}
The grand-canonical free energy \eqref{eq:Landau} fixes the boundary
chemical potential. At fixed density, by contrast, the chemical
potential changes when either condensate forms. Let
$\chi=(\partial\rho_N/\partial\mu)_{T,J}$ be the charge
susceptibility of the normal solution. The first law
\eqref{eq:general_first_law} implies
\begin{equation}
 \chi>0,\qquad
 \left.\frac{\partial s_{\rm th}}{\partial T}\right|_{\rho,J}
 =\left.\frac{\partial s_{\rm th}}{\partial T}\right|_{\mu,J}
 -\frac{1}{\chi}\left(\left.\frac{\partial\rho}{\partial T}\right|_{\mu,J}\right)^2>0
 \label{eq:response}
\end{equation}
for positive normal-state compressibility and fixed-density heat
capacity. These thermal conditions are separate from the stability
of the two condensate amplitudes. They must be checked on the
chosen normal background before interpreting a positive quartic
form as local thermodynamic stability.

Differentiating \eqref{eq:Landau} with respect to $\mu$ gives the
charge induced by the orders,
\begin{equation}
 \rho=\rho_N(T,\mu,J)-\eps\left[
 (\partial_\mu a_\omega)w_0^2
 +(\partial_\mu a_\Phi)s_0^2\right]
 +O(w_0^4,s_0^4).
 \label{eq:density_from_orders}
\end{equation}
To keep $\rho$ at its normal-state value, expand
$\rho_N(T,\mu+\delta\mu,J)=\rho_N(T,\mu,J)+\chi\delta\mu+\cdots$.
Equating this variation to \eqref{eq:density_from_orders} fixes
\begin{equation}
 \delta\mu=\frac{\eps}{\chi}
 \left[(\partial_\mu a_\omega)w_0^2
 +(\partial_\mu a_\Phi)s_0^2\right]
 +O(w_0^4,s_0^4).
 \label{eq:mu_shift}
\end{equation}
Inserting the shift into $F=\Omega+\mu\rho$, the terms linear in
$\delta\mu$ cancel. The fourth-order term that remains is
\begin{equation}
 \frac{F-F_N}{\eps}
 =a_\omega w_0^2+a_\Phi s_0^2
 +b_\omega^{(\rho)}w_0^4+b_\Phi^{(\rho)}s_0^4
 +g^{(\rho)}w_0^2s_0^2+O(6),
 \label{eq:canonical_potential}
\end{equation}
where the fixed-density coefficients follow individually from
\begin{align}
 b_\omega^{(\rho)}&=b_\omega^{(\mu)}+
 \frac{\eps}{2\chi}(\partial_\mu a_\omega)^2,
 &b_\Phi^{(\rho)}&=b_\Phi^{(\mu)}+
 \frac{\eps}{2\chi}(\partial_\mu a_\Phi)^2,
 \nonumber\\
 g^{(\rho)}&=g^{(\mu)}+
 \frac{\eps}{\chi}(\partial_\mu a_\omega)(\partial_\mu a_\Phi).
 \label{eq:canonical}
\end{align}
This rank-one correction is nonnegative when $\chi>0$.
Its effect on mixed-phase stability must nevertheless be considered
alongside the density--order fluctuations below.
For the conformal reference $\partial_\mu a_i=-Q_i$ and
$\chi=\chi_{\rm UV}$ after the common probe normalization is
removed. Equation~\eqref{eq:canonical} then gives the explicitly
integrated fixed-charge coefficients \eqref{eq:canonical_coefficients}.

The density and condensate fluctuations can be examined without
introducing an additional response matrix. Write
$\delta w_0^2=\delta(w_0^2)$ and
$\delta s_0^2=\delta(s_0^2)$ at the normal double onset.
To quadratic order in these variations the Helmholtz free energy is
\begin{align}
 \delta^2F={}&\frac{1}{\chi}
 \left[\delta\rho+\eps(\partial_\mu a_\omega)\delta w_0^2
 +\eps(\partial_\mu a_\Phi)\delta s_0^2\right]^2
 \nonumber\\
 &+\eps\left[2b_\omega^{(\mu)}(\delta w_0^2)^2
 +2g^{(\mu)}\delta w_0^2\delta s_0^2
 +2b_\Phi^{(\mu)}(\delta s_0^2)^2\right].
 \label{eq:relaxed_density_condition}
\end{align}
This expression follows by inserting \eqref{eq:canonical} into
the second variation of \eqref{eq:canonical_potential}, together
with $\partial_\rho^2F_N=1/\chi$ and
$\partial_\rho a_i=(\partial_\mu a_i)/\chi$.
For $\chi>0$, positivity under simultaneous density and order
fluctuations requires precisely the two positive self-couplings and
$4b_\omega^{(\mu)}b_\Phi^{(\mu)}-(g^{(\mu)})^2>0$.
If that last quantity is negative, choosing $\delta\rho$ to
cancel the square leaves a direction in which the free energy
decreases. It is the local phase-separation direction that survives
even when the fixed-density quartic coefficients are positive.

For later use, the determinants of the fixed-$\mu$ and fixed-$\rho$
quartic forms are related by the explicit coefficient identity
\begin{align}
 \det\mathbf B_\rho={}&\det\mathbf B_\mu+\frac{\eps}{\chi}\Big[
 2b_\Phi^{(\mu)}(\partial_\mu a_\omega)^2
 +2b_\omega^{(\mu)}(\partial_\mu a_\Phi)^2
 -2g^{(\mu)}(\partial_\mu a_\omega)
 (\partial_\mu a_\Phi)\Big].
 \label{eq:Schur_identity}
\end{align}
In the ultraviolet reference, the fixed-density threshold can
therefore lie below the fixed-chemical-potential threshold,
as shown in Section~\ref{sec:perturbation}.

At the infrared double onset the quartic coefficients take the form
of the overlap integrals in Eqs.~\eqref{eq:overlap_integrals}
and~\eqref{eq:analytic_all_quartic_terms}
with the $\AdS_8$ measure. Let $x=1/v$, so that the boundary is at
$x=0$ and the horizon at $x=1$, and let $w_1$ and $s_1$ be the nodeless
eigenfunctions of Eqs.~\eqref{eq:IR_SL_p} and~\eqref{eq:IR_SL_s} at
$\widehat\mu=\widehat\mu_p$ and $M=M_*$.
Each condensate sources the Cartan equation at second order in
its amplitude. The resulting electric-potential change acts back
on both condensates, generating quartic interactions even without
a direct scalar--vector contact term. At fixed chemical potential
this response vanishes at the boundary and the horizon.
The electric sources and the corresponding Dirichlet Green
function are
\begin{equation}
 S^{\rm IR}_\omega=\frac{b_0\,w_1^2}{x^4h},\qquad
 S^{\rm IR}_\Phi=\frac{b_0\,s_1^2}{4x^6h},\qquad
 G_{\rm IR}(x,x')=\frac15\,x_<^5\,(1-x_>^5),
 \label{eq:IR_electric_Green}
\end{equation}
with $-\partial_x(x^{-4}\partial_xG_{\rm IR})=\delta(x-x')$ and
$b_0=\mu(1-x^5)$. The kernel obeys
\begin{equation}
 G_{\rm IR}(0,x')=G_{\rm IR}(1,x')=0,\qquad
 \left[-x^{-4}\partial_xG_{\rm IR}(x,x')\right]_{x=x'_-}^{x=x'_+}=1.
 \label{eq:IR_green_jump}
\end{equation}
The two homogeneous solutions are $x^5$ and $1-x^5$.
The first vanishes at the boundary, the second at the horizon,
and their product in \eqref{eq:IR_electric_Green} is continuous
at $x=x'$ with precisely the unit flux jump
\eqref{eq:IR_green_jump}. The overlap and contact integrals are
\begin{equation}
 \begin{gathered}
 I^{\rm IR}_{ij}=\int_0^1\!\!\int_0^1S^{\rm IR}_i(x)\,G_{\rm IR}(x,x')\,S^{\rm IR}_j(x')\,\dd x\,\dd x',\qquad
 Q^{\rm IR}_i=\int_0^1(1-x^5)\,S^{\rm IR}_i\,\dd x,\\
 H^{\rm IR}=\frac18\int_0^1\frac{w_1^2s_1^2}{x^{6}}\,\dd x,\qquad
 A^{\rm IR}_\Phi=\frac14\int_0^1\frac{s_1^4}{x^{8}}\,\dd x,
 \end{gathered}
 \label{eq:IR_direct_integrals}
\end{equation}
For $i,j\in\{\omega,\Phi\}$, the integral $I^{\rm IR}_{ij}$
measures this electric response. Its diagonal entries describe
each condensate acting on itself through the electric field.
The off-diagonal entry describes the field produced by one
condensate acting on the other. The integral $H^{\rm IR}$ is
the direct gauge--Higgs interaction, while $A_\Phi^{\rm IR}$
multiplies the scalar self-interaction. Adding these contributions
gives the grand-canonical coefficients
\begin{equation}
 b_\omega=\frac12I^{\rm IR}_{\omega\omega},\qquad
 b_\Phi=\frac12I^{\rm IR}_{\Phi\Phi}+\lambda_8L_8^2A^{\rm IR}_\Phi,\qquad
 g=I^{\rm IR}_{\omega\Phi}+H^{\rm IR}.
 \label{eq:IR_quartic_coefficients}
\end{equation}
The $\AdS_8$ radius is the infrared unit of length, so the scalar
coupling enters through $\lambda_8L_8^2$. The normal Cartan flux is
$5\mu$ in these units, so the susceptibility is $\chi_{\rm IR}=5$. The
threshold $\lambda_c^{\rm IR,(\sigma)}$ is defined by
Eq.~\eqref{eq:cartan_formal_threshold} with these coefficients, and
Eq.~\eqref{eq:threshold_shift} holds with $\chi=\chi_{\rm IR}$. Each
threshold is independent of the normalizations of $w_1$ and $s_1$.
These coefficients belong to the infrared endpoint. On the
complete flow, the profiles must instead satisfy the ultraviolet
source conditions, and their coupled electric, scalar and metric
responses enter the quartic action.

At an infrared double onset, positivity of the quartic form
requires $\lambda_8L_8^2>\lambda_c^{\rm IR,(\mu)}$.
The mixed squared amplitudes must also be positive for the
stationary state to be physical. In the conformal reference,
$\lambda_8L_5^2<\lambda_c^{\rm UV,(n)}$ makes both endpoint
quartic forms indefinite. These different endpoint tests can
therefore be separated by a parent coupling in the interval
\begin{equation}
 \boxed{\;\frac{\lambda_c^{\rm IR,(\mu)}}{L_8^2}<\lambda_8<
 \frac{\lambda_c^{\rm UV,(n)}}{L_5^2}\;}
 \label{eq:window}
\end{equation}
This interval is nonempty when its lower endpoint is smaller
than its upper endpoint. The vacuum relations
\eqref{eq:vacuum_relations} give
$L_8^2/L_5^2=28/(8-\lambda_\Psi f_{\rm UV}^4L_5^2)$,
so the parent winding data affect the comparison of the two
dimensionful thresholds. The ultraviolet coefficient is given by
exact radial profiles, while the infrared threshold is defined
by eigenfunction overlap integrals. Their comparison determines
whether the interval \eqref{eq:window} is open.
Equation~\eqref{eq:threshold_shift} orders the ensembles within
each endpoint. The common onset and mixed-phase stability on the
interpolating geometry require Eqs.~\eqref{eq:critical}
and~\eqref{eq:quartic} with full radial profiles.

\subsection{Higher-dimensional interpretation and stability bounds}
\label{sec:parent_checks}
\paragraph{Nonlinear free-energy comparison.}
The quartic test concerns small condensates. At finite amplitude,
the charge carried by each solution determines how its free
energy changes with chemical potential.
Within the torus ansatz \eqref{eq:uplift}, any solution of the retained
five-dimensional equations uplifts to a saddle of the same
eight-dimensional Einstein--matter theory. The quartic analysis
therefore compares ordered saddles within one parent theory, but
only near the charged onset. To compare them beyond the small-amplitude
region, one must retain the full radial responses and impose the
same boundary temperature, chemical potential, metric and neutral
source. The Cartan equation \eqref{eq:full_cartan_flux} expresses the
boundary density in terms of the electric flux at any regular
radial point $r_a$ and the charge carried by the condensates,
\begin{equation}
 \rho=\eps\lim_{R\to\infty}\left[
 \left.\frac{e^{2A+B}}Xb'\right|_{r_a}
 +\int_{r_a}^{R}\dd r\,\frac b h
 \left(\frac{e^{-B}}Xw^2+\frac{e^{2A+B}}4s^2\right)\right].
 \label{eq:charge}
\end{equation}
The right-hand side is independent of $r_a$. For two nonlinear
solutions evaluated at the same fixed $T$ and $J$, the first law
gives
\begin{equation}
 \dd(\Omega_a-\Omega_b)=-(\rho_a-\rho_b)\,\dd\mu .
 \label{eq:firstlaw}
\end{equation}
For example, if the mixed branch joins the pure vector branch
continuously at $\mu_{pM}$, its free-energy difference is
\begin{equation}
 \Omega_{s+p}(T,\mu)-\Omega_p(T,\mu)
 =-\int_{\mu_{pM}}^\mu
 [\rho_{s+p}(T,\mu')-\rho_p(T,\mu')]\,\dd\mu'.
 \label{eq:nonlinear_free_energy}
\end{equation}
The scalar branch gives the analogous relation. Evaluating these
integrals along nonlinear branches at fixed $(T,J)$ extends the
near-onset comparison \eqref{eq:analytic_gaps} to finite amplitudes.
The equivalent on-shell-action prescription is
$\Omega=TI_E^{\rm ren}/V_3$, with the $\mathcal F K$ boundary term
required by \eqref{eq:S5} and the BF scalar counterterm.
Such a comparison would cover homogeneous normal, pure and mixed
black-brane solutions in the retained ansatz, as in
Ref.~\cite{Nie:2014backreaction}. It would not rank
inhomogeneous saddles, compactification caps or internally polarized
Yang--Mills configurations.

\paragraph{Transverse gravitational fluctuations.}
The ordered background distinguishes $x_1$ but preserves rotations
in the $x_2,x_3$ plane. Its transverse $x_2x_3$ shear therefore
decouples from scalar and vector fluctuations. Write its
dimensionless radial amplitude as $H_{23}(r)$, with frequency
$\omega$, external momentum $q$ along $x_1$, and integer torus
momentum $\mathbf n$. The torus wave number is $k$,
\begin{equation}
 \delta\widehat g_{x_2x_3}=Xe^{2A}H_{23}(r)e^{-i\omega t+iqx_1+ik\mathbf n\cdot\mathbf y},
 \qquad\mathbf n\in\mathbb Z^3.
\end{equation}
Here $A$ is the external warp factor, $B$ describes the spatial
anisotropy, $h$ is the blackening function and $X$ sets the torus
size in Eq.~\eqref{eq:uplift}. The curvature coefficient
$\mathcal F=1-\eps\xi s^2/2$ must be positive for a positive
gravitational kinetic term. The radial equation balances the
radial, external-momentum and internal-momentum contributions
against the squared frequency,
\begin{equation}
 -(\mathcal Fe^{4A+B}hH_{23}')'
 +\mathcal F\left(e^{2A-B}q^2+e^{4A+B}X^2k^2|\mathbf n|^2\right)H_{23}
 =\omega^2\frac{\mathcal Fe^{2A+B}}hH_{23} .
 \label{eq:TT}
\end{equation}
A mode with $\operatorname{Im}\omega>0$ would grow in time
because its time dependence is $e^{-i\omega t}$.
At a nonextremal horizon the ingoing profile is
$H_{23}\sim(r-r_h)^{-i\omega/(4\pi T)}$.
For such a growing mode it vanishes at the horizon, the radial
integrals converge, and the integration-by-parts boundary term
is zero. The source-free ultraviolet falloff removes the other
boundary term. Multiplying by $H_{23}^*$ and integrating gives
\begin{multline}
 \int_{r_h}^\infty\!\dd r\,\mathcal F\Big[e^{4A+B}h|H_{23}'|^2
 +\big(e^{2A-B}q^2+e^{4A+B}X^2k^2|\mathbf n|^2\big)|H_{23}|^2\Big]
 \\
 =\omega^2\int_{r_h}^\infty\!\dd r\,\frac{\mathcal Fe^{2A+B}}{h}|H_{23}|^2 .
 \label{eq:TT_identity}
\end{multline}
The left-hand side is real and nonnegative, and the integral on the right is positive. Hence $\omega^2$ is real and nonnegative. For a frequency in the upper half-plane, reality first requires $\operatorname{Re}\omega=0$, after which $\omega^2<0$ is a contradiction. Thus no growing mode exists in this channel, for every torus harmonic, provided $\mathcal F,h>0$ outside the horizon. This argument includes the nonminimal coupling and the anisotropic ordered background.

\paragraph{Winding scalar fluctuations.}
For this separate test, hold the metric, charged doublet and
Yang--Mills perturbations fixed. Factoring out the background
phase $e^{iky^I}$ shifts the internal momentum by one winding
unit. The complex-conjugate component has the opposite shift.
The two momentum eigenvalues are
\begin{equation}
 X^2k^2\big[(n_I\pm1)^2+|\mathbf n_\perp|^2\big]\ge0 .
\end{equation}
Here $\mathbf n_\perp$ contains the two components orthogonal
to the winding direction $I$. Remove the radial measure by
writing the scalar fluctuation amplitude as $\delta f_{I,\mathbf n}$,
\begin{equation}
 \delta\Psi_I=e^{iky^I+ik\mathbf n\cdot\mathbf y}
 e^{-2A-B/2}\delta f_{I,\mathbf n}(r).
\end{equation}
With an overall positive normalization suppressed,
$Q_{\mathbf n}$ is the static second variation of the winding
action at zero external momentum. Its squared-frequency problem
has the positive norm
$\int_{r_h}^\infty e^{-2A}h^{-1}|\delta f_{I,\mathbf n}|^2\,\dd r$.
The radial kinetic term becomes
$h|\delta f_{I,\mathbf n}'-(2A'+B'/2)\delta f_{I,\mathbf n}|^2$.
Expanding this square, integrating its cross term by parts,
and dropping the nonnegative internal-momentum contribution gives
\begin{align}
 Q_{\mathbf n}\ge\int_{r_h}^\infty\dd r\,\bigg\{
 &h\big|\partial_r\delta f_{I,\mathbf n}\big|^2
 +\bigg[h\left(2A'+\frac{B'}2\right)^2
 +\left[h\left(2A'+\frac{B'}2\right)\right]'\nonumber\\
 &\quad+X\left(m_\Psi^2+\lambda_\Psi f^2
 +\frac{\eps\eta}{2}s^2\right)\bigg]
 |\delta f_{I,\mathbf n}|^2\bigg\}.
 \label{eq:bound}
\end{align}
For $\lambda_\Psi\ge0$ the additional splitting between the
neutral amplitude modes contributes a nonnegative term, while the
portal coupling retains the sign of $\eta$. A sufficient condition
for the fixed-background winding sector to have no negative mode is
therefore
\begin{align}
 h\left(2A'+\frac{B'}2\right)^2
 +\left[h\left(2A'+\frac{B'}2\right)\right]'
 +X\left(m_\Psi^2+\lambda_\Psi f^2
 +\frac{\eps\eta}{2}s^2\right)\ge0,
 \quad r\ge r_h .
 \label{eq:frozen_sufficient}
\end{align}
With regular horizon and normalizable ultraviolet boundary
conditions, Eq.~\eqref{eq:frozen_sufficient} gives stability
of the winding fluctuations at fixed metric and charged fields.
The coupled spectrum is determined by varying all these fields
together.

\section{Discussion}\label{sec:discussion}
The principal geometric observation is that the ultraviolet and
infrared scaling regimes can be described within a single classical
eight-dimensional Einstein--matter theory. Three neutral scalars
with winding phases support the product $\AdS_5\times T^3$ vacuum,
while their vanishing amplitudes permit a locally $\AdS_8$
region. This construction also fixes the interactions used to study
charged order. The potential \eqref{eq:potential}, the gauge factor
$Z=X^{-1}$, the charged effective mass and the scalar self-coupling
follow from the same reduction. In particular,
$Z\lambda_\Phi=\lambda_8$ at every radius, and the vacuum relations
\eqref{eq:vacuum_relations} tie the curvature and winding scales
together. The model therefore does not allow its infrared charged
couplings to be chosen independently of the ultraviolet theory.
This restriction is essential when the aim is to compare different
orders along one radial evolution rather than to assemble separate
effective models at its endpoints.

The change in the scaling geometry has a precise, and limited,
physical meaning. At the ultraviolet boundary the field theory is
four-dimensional, with sources and expectation values defined using
the $\AdS_5$ holographic dictionary. In the infrared the winding
amplitude has the exponential decay \eqref{eq:winding_Bessel}, so
the external and internal directions share the local $\AdS_8$
scale factor. Reducing this local metric gives $z=1$ and
$\theta=-3$. The difference appears, for example, in the neutral
entropy scaling $T^3$ versus $T^6$ and in the zero-density optical
power $\omega$ versus $\omega^4$. These are signatures of a
higher-dimensional local geometry, not a claim that the ultraviolet
boundary theory turns into a seven-dimensional conformal field
theory. The torus remains periodically identified, and the
five-dimensional description retains the same boundary spacetime.
The complete crossover functions still depend on the solution of
the neutral radial problem \eqref{eq:neutral_scalars}--\eqref{eq:neutral_einstein}.

The parent construction also restricts which ultraviolet
deformations and charged instabilities are admissible. The two
neutral singlet masses follow from one quadratic mass matrix, rather than from
independent mass assignments. The interval \eqref{eq:singlet_window}
selects one relevant neutral deformation while respecting the
$\AdS_5$ BF bound, and its source $J$ is held fixed when the
charged saddles are compared. The charged scalar saturates the
ultraviolet BF bound by the parameter relation
\eqref{eq:parameters}. Its infrared mass instead receives the
parent curvature coupling and must obey
\eqref{eq:charged_IR_BF}. Along an ordered solution, positivity
of the effective Einstein coefficient $\mathcal F$ is another
independent restriction. These conditions prevent a change of
boundary source, a violation of the BF bound or a singular
gravitational coupling from being mistaken for a transition
between legitimate superfluid states.

The distinction between a linear instability and a thermodynamic
phase is particularly important for competing orders. The exact
conformal zero modes \eqref{eq:zero_modes} make the ultraviolet
quartic terms accessible by radial integration, including the
response of the electric potential. In the infrared, the scalar
and vector Sturm--Liouville problems
\eqref{eq:IR_SL_p}--\eqref{eq:IR_SL_s} determine their separate
onset scales. The radial estimates in Section~\ref{sec:phase}
give a unique coincidence mass $M_*$ above the infrared BF bound,
realized by the fixed parent couplings
\eqref{eq:IR_parent_mass_matching}. On the complete flow, both
charged modes must satisfy the ultraviolet source conditions.
Their common onset is specified by Eq.~\eqref{eq:critical}, with
independent slopes as $T$ and $\mu$ vary at fixed $J$.

When the two orders are marginal together, the sign of a single
quartic coefficient is not enough to decide coexistence. The
matrix \eqref{eq:quartic} must be positive on the two independent
condensate directions, and the mixed stationary amplitudes must
both be real and nonzero. Under these hypotheses, the free-energy
identities \eqref{eq:analytic_gaps} show that the mixed state lies
below both pure states sufficiently close to the common onset.
The four continuous boundaries then have the tetracritical
arrangement shown in Fig.~\ref{fig:phase}. Conversely, a negative
quartic determinant produces a direction in which the mixed
state can lower its free energy by changing the relative
condensates. These are analytic implications of the local
functional, not a claim that the full-flow criteria have already
been met. In the latter problem the quartic coefficients also
receive contributions from changes in the metric and neutral
supporting fields.

An additional distinction arises between fixed chemical potential
and fixed Cartan charge density. Forming either condensate changes
the electric flux, so the chemical potential must relax when the
total density is constrained. This produces the explicit
Legendre-transform correction \eqref{eq:canonical} and the
determinant relation \eqref{eq:Schur_identity}. The threshold
ordering \eqref{eq:threshold_shift} shows that a positive
fixed-density quartic form can arise before the corresponding
fixed-potential form becomes positive as $\lambda_8$ increases.
Nevertheless, the joint density--order variation
\eqref{eq:relaxed_density_condition} still contains an unstable
direction when the grand-canonical determinant is negative.
In that intermediate regime, a homogeneous fixed-density
extremum should not be identified with a stable mixed phase.
allowing density redistribution reveals the local tendency to
phase separation. This distinction is pertinent to holographic
models of systems in which the charge is controlled globally
rather than through a reservoir.

The endpoint thresholds connect this competition directly to the
parent geometry. The ultraviolet integrals follow from closed-form
zero modes. The infrared coefficients instead involve overlaps
of $\AdS_8$ eigenfunctions with the electric Green function.
The two thresholds must be expressed in the same eight-dimensional
coupling convention before they are compared, which accounts for
the factors of $L_5^2$ and $L_8^2$ in Eq.~\eqref{eq:window}.
That interval identifies a possible contrast between the two
endpoint quartic forms when its ordering can be established.
It does not imply that the quartic matrix changes sign at a
particular point on the complete radial flow. The actual onset
and quartic response depend on the full support of the charged
wavefunctions and of the metric and neutral variations between
the endpoints.

This separation matters even when an endpoint calculation
suggests a favorable mixed order. The scalar quartic interaction
enters the onset functional linearly through a positive profile
integral, as expressed in Eq.~\eqref{eq:affine_quartic}. The
result gives a definite analytic way to locate the threshold
once the background modes are known. It does not determine the
complete-flow critical coupling without those modes. The
same caution applies to the canonical ensemble. Its quartic
matrix may be positive even when relaxation of the local
charge density reveals an unstable direction. Thus a phase
diagram based only on endpoint modes or on one constrained
thermodynamic ensemble can be misleading.

Beyond the near-onset regime, a matched thermodynamic comparison
requires tracing the nonlinear branches at identical $(T,\mu,J)$.
The charge relation \eqref{eq:charge} separates the horizon flux
from the charge carried by condensates. The first-law integral
\eqref{eq:nonlinear_free_energy} would then compare the mixed and
pure free energies from a common branch-joining point. A direct
evaluation of that integral or the matched renormalized on-shell
actions is distinct from the analytic quartic ordering close to
a hypothetical common onset. The present endpoint calculations
also cannot rule out first-order transitions or lower-action
branches elsewhere in the phase diagram.

Finally, consistency of the background truncation does not imply
stability under every fluctuation of the eight-dimensional theory.
The transverse spin-two identity \eqref{eq:TT_identity} excludes
growing modes in that decoupled channel, including its torus
harmonics. The winding-scalar condition
\eqref{eq:frozen_sufficient} is sufficient only when metric and
charged fluctuations are held fixed. In addition, an internal
Yang--Mills polarization $B_I^1$ is a homogeneous field of the
parent theory that is absent from the retained ordered ansatz.
Its possible competition with an external vector condensate
cannot be decided by a calculation restricted to the latter.
The complete coupled Kaluza--Klein spectrum and a nonlinear
comparison including internal vector branches are therefore
separate requirements for global stability. This is consistent
with the role of fluctuation tests in generalized reductions
\cite{GouterauxSmolic:2012,Gath:2013,CremoniniSinkovics:2014,GKN:2026}.

The results separate three levels of information. The
eight-dimensional action fixes a five-dimensional matter sector
and the local ultraviolet and infrared scaling solutions. Exact
ultraviolet modes and analytic infrared spectral relations give
the corresponding onset and quartic tests. Finally, the
complete-flow criteria specify what would be needed to establish
a mixed phase with a definite free-energy preference. The
existence of a transverse double onset and a positive quartic
matrix for one complete interpolating solution has not been
established by the analytic endpoint calculations presented here.
Neither the reduction nor these local criteria establish a
string embedding, global saddle preference or the stability of
the complete coupled Kaluza--Klein spectrum.

\appendix

\section{Toroidal reduction and exact ultraviolet data}
\label{app:general_reduction}
This appendix collects intermediate steps in the reduction of
Section~\ref{sec:reduction} and the conformal reference calculation
of Section~\ref{sec:charged}. We begin with the metric and matter
densities on a torus in $D=d+n$ dimensions, before specializing
to the vacuum fluctuations and charged perturbations. This sets
the conventions for comparison with generalized dimensional
reduction \cite{GouterauxSmolic:2012,Smolic:2013,GKN:2026}.

Let $\widehat g_{MN}$ be a $D$-dimensional metric and $g_{\mu\nu}$
the external $d$-dimensional Einstein-frame metric. For a flat
internal torus and a single canonically normalized volume modulus,
take
\begin{equation}
 \dd s_D^2=e^{2\alpha\varphi}\dd s_d^2
 +e^{2\beta\varphi}\delta_{IJ}\dd y^I\dd y^J,
 \quad
 \alpha^2=\frac{n}{2(d-2)(D-2)},
 \qquad \beta=-\frac{d-2}{n}\alpha.
 \label{eq:general_KK_ansatz}
\end{equation}
The positive sign of $\alpha$ is a convention for $\varphi$.
The Einstein-frame normalization fixes only its magnitude and the
ratio $\beta/\alpha$. The metric determinant and scalar curvature
then satisfy
\begin{align}
 \sqrt{-\widehat g}
 &=\sqrt{-g}\,e^{(d\alpha+n\beta)\varphi}
 =\sqrt{-g}\,e^{2\alpha\varphi},\nonumber\\
 \widehat R
 &=e^{-2\alpha\varphi}\left[
 R_d-\frac12(\partial\varphi)^2
 -2\alpha\Box_d\varphi\right].
 \label{eq:general_ricci}
\end{align}
In the absence of a field-dependent coefficient of $\widehat R$,
the last term is a boundary derivative and the reduced scalar
kinetic term is canonical. No Weyl rescaling is left to be performed
after using \eqref{eq:general_KK_ansatz}.

Consider an external Yang--Mills field, an external complex scalar
and a scalar potential $\widehat U$. Their reduced densities obey
\begin{align}
 \sqrt{-\widehat g}\,\widehat G_{\mu\nu}
 \widehat G^{\mu\nu}
 &=\sqrt{-g}\,e^{-2\alpha\varphi}
 G_{\mu\nu}G^{\mu\nu},\nonumber\\
 \sqrt{-\widehat g}\,|\widehat D_\mu\Phi|^2
 &=\sqrt{-g}\,|D_\mu\Phi|^2,\nonumber\\
 \sqrt{-\widehat g}\,\widehat U(|\Phi|)
 &=\sqrt{-g}\,e^{2\alpha\varphi}
 \widehat U(|\Phi|).
 \label{eq:general_matter_weights}
\end{align}
An internal gradient contributes instead
$e^{2(\alpha-\beta)\varphi}|\partial_I\Psi|^2$.
These relations explain the opposite running of the Yang--Mills
kinetic function and the charged scalar potential, independent of
the particular neutral profile. If a term $-\xi\widehat R\,\boldsymbol\Phi^\dagger\boldsymbol\Phi$
is present in $D$ dimensions, it must be reduced \emph{before}
partial integration and variation. Its radion derivative mixing
is proportional to $\partial_\mu\mathcal F\partial^\mu\varphi$ and survives
in the fully backreacted equations.

For the fixed specialization $d=5$, $n=3$, write
$X=e^{\varphi/\sqrt3}$. The ansatz has
$e^{2\alpha\varphi}=X$ and $e^{2\beta\varphi}=X^{-1}$.
Three windings $\Psi_I=f e^{iky^I}$ contribute the exact sum
\begin{equation}
 \sum_I\sqrt{-\widehat g}\,|\partial_I\Psi_I|^2
 =\sqrt{-g}\,3k^2X^2f^2.
 \label{eq:winding_reduction}
\end{equation}
The remaining mass and quartic terms are multiplied by $X$.
Combining them with the cosmological constant gives precisely
Eq.~\eqref{eq:potential}. The action-level identity
$Z\lambda_\Phi=\lambda_8$ is therefore not specific to a
particular infrared geometry. It reflects the complementary
metric weights in \eqref{eq:general_matter_weights}.

\subsection{\texorpdfstring{Vacua and the singlet BF test}{Vacua and the singlet BF test}}
The ultraviolet product vacuum solves
$\partial_\phi V=\partial_f V=0$ at $X=1$ and
$f=f_{\rm UV}$. The scalar equation and the internal Einstein
equation reduce to the first two conditions in
\eqref{eq:vacuum_relations}. Eliminating $m_\Psi^2$ and the parent
cosmological constant from the value of the potential gives
\begin{equation}
 V(1,f_{\rm UV})=-3k^2f_{\rm UV}^2
 =-\frac{12}{L_5^2},\qquad
 L_5^2=\frac4{k^2f_{\rm UV}^2}.
 \label{eq:symbolic_UV_vacuum}
\end{equation}
The last equality explains the third relation in
\eqref{eq:vacuum_relations} when $L_5=1$. It expresses the
external radius through the winding data.

The charged BF constraint follows by evaluating the parent
curvature coupling at this product vacuum.
\begin{equation}
 m_{\Phi,\rm UV}^2=m_0^2+3\eta f_{\rm UV}^2+\xi R_{8,\rm UV},
 \qquad R_{8,\rm UV}=-\frac{20}{L_5^2}.
 \label{eq:UV_BF_from8}
\end{equation}
Direct differentiation with respect to the two canonically
normalized neutral amplitudes gives the quadratic form
\eqref{eq:BF_matrix_test}. Its two eigenvalues are
\eqref{eq:singlet_masses}, and their characteristic equation is
\begin{equation}
 (t-m_+^2)(t-m_-^2)
 =t^2-2(k^2+\lambda_\Psi)f_{\rm UV}^2t
 +4k^2\lambda_\Psi f_{\rm UV}^4-2k^4f_{\rm UV}^2.
 \label{eq:BF_characteristic}
\end{equation}
With $k^2f_{\rm UV}^2=4$, the strict BF condition that both
roots lie above $-4/L_5^2=-4$ becomes
\begin{equation}
 16+2\lambda_\Psi f_{\rm UV}^2>0,\qquad
 6+3\lambda_\Psi f_{\rm UV}^2-k^2>0 .
 \label{eq:UV_polynomial_BF}
\end{equation}
The first inequality tests the shifted sum of eigenvalues and the
second their shifted product. Together with the relevance
condition $2\lambda_\Psi f_{\rm UV}^2<k^2$ they give
Eq.~\eqref{eq:singlet_window}, inside which the trace condition holds
automatically. Relative-amplitude and torus-shape masses belong to the
full eight-dimensional fluctuation system of
Section~\ref{sec:parent_checks}.

The locally $\AdS_8$ endpoint occurs where the winding amplitude
vanishes and the torus acquires the same local scale factor as the
external geometry. Its radius is fixed by the cosmological
constant in the eight-dimensional action. The corresponding
five-dimensional scaling solution has $\theta=-n$ and $z=1$.
These local relations follow from an eight-dimensional solution
with an integer number of internal dimensions.

\subsection{The background consistency test}
The parent Einstein and scalar equations are given by
\eqref{eq:8dEinstein} and \eqref{eq:parent_scalar_equations} in
Section~\ref{sec:reduction}. They are used here without introducing
a second set of gravitational field equations. The reduction of the
mixed and traceless internal projections follows from
\eqref{eq:internal_stress_consistency}--\eqref{eq:omitted_equations}.
The three neutral scalar amplitudes satisfy the same radial
equation, while the sum of the internal Einstein components is
precisely the radion equation obtained by varying \eqref{eq:S5}.
In this way, any solution of the retained five-dimensional
equations uplifts to the eight-dimensional background ansatz.
The omitted shape and internal-vector fluctuations still have
independent linearized equations, as discussed in
Section~\ref{sec:parent_checks}.

\subsection{Exact ultraviolet reference calculations}\label{app:analytic_proof}
\subsubsection{Exact AdS cubic profiles and quartic integrals}\label{app:cubic}
The following exact calculations refer to the
$\AdS_5\times T^3$ conformal background. They are independent
of the neutral two-field evolution and supply analytically
normalized boundary conditions for its charged modes.

At cubic order, the vector and scalar equations inherit the radial
derivatives and potentials of Eqs.~\eqref{eq:vector_zero_equation}
and~\eqref{eq:scalar_zero_equation},
\begin{align}
 (u^3f_u\omega_3')'+\frac{b_c^2(1-u^{-2})^2}{uf_u}\omega_3
 &=R_\omega,\nonumber\\
 (u^5f_u\Phi_3')'+4u^3\Phi_3
 +\frac{b_c^2u(1-u^{-2})^2}{4f_u}\Phi_3
 &=R_\Phi.
\end{align}
Their sources are determined directly by the electric profiles in
Eqs.~\eqref{eq:b2I_solution}--\eqref{eq:b2II_solution},
\begin{align}
 R_\omega={}&-\frac{8w_0(1-u^{-2})\psi}{uf_u}
 \left[\delta b(1-u^{-2})+w_0^2b_2^\omega+s_0^2b_2^\Phi\right]
 +\frac{w_0s_0^2}{4}u\Phi_1^2\psi,
 \label{eq:omega3_full}\\
 R_\Phi={}&-\frac{2s_0 u(1-u^{-2})\Phi_1}{f_u}
 \left[\delta b(1-u^{-2})+w_0^2b_2^\omega+s_0^2b_2^\Phi\right]
 \nonumber\\
 &+\frac{w_0^2s_0}{4}u\psi^2\Phi_1
 +\lambda_8s_0^3u^3\Phi_1^3.
 \label{eq:Phi3_full}
\end{align}
Multiplication by a zero mode converts the left-hand side to a total
derivative. Horizon regularity and the source-free boundary conditions
then give
\begin{equation}
 \int_1^\infty\psi R_\omega\,\dd u=0,\qquad
 \int_1^\infty\Phi_1 R_\Phi\,\dd u=0.
 \label{eq:cubic_solvability}
\end{equation}
These two conditions reproduce Eqs.~\eqref{eq:vector_amplitude} and
\eqref{eq:scalar_amplitude}. With no additional normalizable coefficient,
the regular cubic solutions are
\begin{align}
 \omega_3(u)&=-\psi(u)\int_u^\infty\frac{\dd v}{v^3f_u(v)\psi(v)^2}
 \int_1^v\psi(s)R_\omega(s)\,\dd s,
 \label{eq:omega3_integral}\\
 \Phi_3(u)&=-\Phi_1(u)\int_u^\infty\frac{\dd v}{v^5f_u(v)\Phi_1(v)^2}
 \int_1^v\Phi_1(s)R_\Phi(s)\,\dd s.
 \label{eq:Phi3_integral}
\end{align}
The zero total integrals in Eq.~\eqref{eq:cubic_solvability} ensure convergence
and remove the scalar logarithmic source. Thus the amplitude equations are
also sufficient for regular perturbative solutions with the prescribed
normalizable amplitudes.

The ultraviolet quartic coefficients are expressed by the
response and contact integrals \eqref{eq:overlap_integrals}. The radial functions in
Eqs.~\eqref{eq:zero_modes} and \eqref{eq:b2I_solution}--
\eqref{eq:b2II_solution} are rational in $u$, so all these integrals
can be evaluated by rational integration after $x=u^{-2}$.
They enter the grand-canonical determinant as
\begin{equation}
 \det\mathbf B_{\rm UV}^{(\mu)}(\lambda_8)=
 4\left(\frac{I_{\omega\omega}}2\right)
 \left(\frac{I_{\Phi\Phi}}2+\lambda_8 A_\Phi\right)
 -\left(I_{\omega\Phi}+H\right)^2.
 \label{eq:UV_symbolic_det}
\end{equation}
The canonical determinant is obtained from the rank-one update
\eqref{eq:canonical_coefficients}. In either ensemble the
coefficient of $\lambda_8$ is strictly positive, and its zero is
the ultraviolet threshold of Eq.~\eqref{eq:cartan_formal_threshold}.

\section{Radial electric response and canonical variation}
\label{app:gravitational_quartics}
At a common normal-state onset, the Cartan potential and the
neutral and metric fields respond at second order in the charged
amplitudes. Here we give the electric equations on the actual
interpolating background. The canonical calculation follows below.

\subsection{Electric response on the normal branch}
Let $(A_N,h_N,\phi_N,f_N,b_N)$ denote a normal solution of
Eq.~\eqref{eq:S5}. The normal branch has $B_N=0$. Its regular,
source-free charged zero modes are $w_\star$ and $s_\star$.
In addition to Eq.~\eqref{eq:metric_response_expansion}, write
\begin{align}
 A&=A_N+w_0^2\delta A_\omega+s_0^2\delta A_s+\cdots,&
 h&=h_N+w_0^2\delta h_\omega+s_0^2\delta h_s+\cdots,
 \nonumber\\
 B&=w_0^2\delta B_\omega+s_0^2\delta B_s+\cdots,&
 \phi&=\phi_N+w_0^2\delta\phi_\omega+s_0^2\delta\phi_s+\cdots,
 \nonumber\\
 f&=f_N+w_0^2\delta f_\omega+s_0^2\delta f_s+\cdots.
 \label{eq:supporting_field_expansion_app}
\end{align}
With $X_N=e^{\phi_N/\sqrt3}$, expansion of the Cartan
equation \eqref{eq:full_cartan_flux} gives the two explicit
second-order equations
\begin{align}
 &\left[\frac{e^{2A_N}}{X_N}\left(
 \delta b_\omega'+\left(2\delta A_\omega+\delta B_\omega
 -\frac{\delta\phi_\omega}{\sqrt3}\right)b_N'\right)\right]'
 =\frac{b_Nw_\star^2}{X_Nh_N},
 \label{eq:charged_electric_response_p}\\
 &\left[\frac{e^{2A_N}}{X_N}\left(
 \delta b_s'+\left(2\delta A_s+\delta B_s
 -\frac{\delta\phi_s}{\sqrt3}\right)b_N'\right)\right]'
 =\frac{e^{2A_N}b_Ns_\star^2}{4h_N}.
 \label{eq:charged_electric_response_s}
\end{align}
The remaining variations of $A,h,B,\phi,f$ obey the Einstein and
neutral equations obtained from \eqref{eq:S5}, together with the
radial constraint. We choose the horizon at a fixed radial
coordinate, impose $\delta b_\omega=\delta b_s=0$ there and at
the boundary at fixed $\mu$, and hold $T$, the boundary metric and
the neutral source $J$ fixed. The metric and neutral variations
therefore contribute to the full-flow quartic response. The
quartic coefficients must be constructed from the on-shell
action including these variations. In the probe limit their terms
drop out of Eqs.~\eqref{eq:charged_electric_response_p}--
\eqref{eq:charged_electric_response_s}, giving the ultraviolet
electric profiles and infrared radial integral expressions of
Sections~\ref{sec:perturbation} and~\ref{sec:phase}.

\subsection{Fixed-density variation of charged condensates}
\label{app:canonical_hessian}
The canonical ensemble fixes the Cartan density, rather than the
chemical potential of the normal solution. This difference can be
derived directly from the expansion \eqref{eq:Landau}. At a fixed
temperature, expand the normal grand potential about $\mu_N$,
where $\rho=-\partial_\mu\Omega_N(\mu_N)$,
\begin{equation}
 \Omega_N(\mu_N+\delta\mu)
 =\Omega_N(\mu_N)-\rho\,\delta\mu
 -\frac{\chi}{2}(\delta\mu)^2+O((\delta\mu)^3),
 \qquad \chi=\left.\frac{\partial\rho_N}{\partial\mu}\right|_{T,J}.
 \label{eq:canonical_expanded_normal}
\end{equation}
For fixed amplitudes, the quadratic coefficients of the charged
orders change with the chemical potential according to
\begin{equation}
 a_\omega(\mu_N+\delta\mu)
 =a_\omega(\mu_N)+(\partial_\mu a_\omega)\delta\mu+\cdots,
 \qquad
 a_\Phi(\mu_N+\delta\mu)
 =a_\Phi(\mu_N)+(\partial_\mu a_\Phi)\delta\mu+\cdots.
 \label{eq:canonical_expanded_orders}
\end{equation}
The electric boundary condition determines the chemical-potential
shift through Eq.~\eqref{eq:mu_shift}. At fixed $\rho$, add
$(\mu_N+\delta\mu)\rho$ to the expanded grand potential.
The linear normal-state contribution cancels, and the terms
containing the shift are exactly
\begin{align}
 &-\frac{\chi}{2}(\delta\mu)^2
 +\eps\delta\mu\left[
 (\partial_\mu a_\omega)w_0^2
 +(\partial_\mu a_\Phi)s_0^2\right]
 \nonumber\\
 &\hspace{2cm}=\frac{\eps^2}{2\chi}
 \left[(\partial_\mu a_\omega)w_0^2
 +(\partial_\mu a_\Phi)s_0^2\right]^2.
 \label{eq:canonical_quartic_from_variation}
\end{align}
Expanding the square gives each term in
Eq.~\eqref{eq:canonical}. In particular, the direct scalar--vector
contact term is not modified, while the induced electric response
contributes to both self-interactions and the mixed quartic.
The shift in $\mu$ is determined independently on each branch,
so two saddles at the same charge density need not have the same
chemical potential.

When local density fluctuations are also admitted, expand the
normal Helmholtz potential and the quadratic charged coefficients,
\begin{align}
 F_N(\rho+\delta\rho)
 &=F_N(\rho)+\mu_N\delta\rho
 +\frac{(\delta\rho)^2}{2\chi}+\cdots,
 \nonumber\\
 a_i(\rho+\delta\rho)
 &=a_i(\rho)+\frac{\partial_\mu a_i}{\chi}
 \delta\rho+\cdots,
 \qquad i=\omega,\Phi.
 \label{eq:density_order_direct_expansion}
\end{align}
Inserting the fixed-density quartic coefficients
\eqref{eq:canonical} into the second variation produces the
completed square in \eqref{eq:relaxed_density_condition}.
One can therefore vary $\delta\rho$ to cancel the first square.
The remaining free-energy variation depends on the original
fixed-chemical-potential coefficients, which must satisfy
$b_\omega^{(\mu)}>0$, $b_\Phi^{(\mu)}>0$ and
$4b_\omega^{(\mu)}b_\Phi^{(\mu)}-(g^{(\mu)})^2>0$.
This density--order result applies locally near a nondegenerate
double onset. The fixed-density heat capacity follows separately
from \eqref{eq:response}, and time-dependent perturbations are
examined in Section~\ref{sec:parent_checks}.

\section{Radial response and coupling thresholds}
\label{app:radial_response}
The endpoint analysis and perturbative phase construction require
radial information beyond the leading background solution.
We collect the neutral current fluctuation equation, the
scalar-coupling dependence of the quartic functional, and the
boundary conditions used in nonlinear radial continuation.

\subsection{Neutral optical response}
In the neutral isotropic background the external
transverse Maxwell perturbation obeys
\begin{equation}
 \partial_r\left(Ze^{2A_N}h_N\partial_ra_x\right)
 +\omega^2\frac{Z}{h_N}a_x=0.
 \label{eq:optical_neutral_ode}
\end{equation}
In the locally higher-dimensional infrared region
$e^{A_N}\propto r^2$, $X\propto r^{-2}$ and $Z=X^{-1}$.
The zero-temperature equation reduces, up to a fixed normalization
of the radial coordinate, to
\begin{equation}
 \partial_r\left(r^{d_s+n}a_x'\right)
 +\omega^2r^{d_s+n-4}a_x=0,
 \label{eq:optical_dimension_count}
\end{equation}
Restoring $L_8$, the ingoing
solution of the parent $\AdS_8$ Maxwell equation is
\begin{equation}
 a_x(r)\propto r^{-(d_s+n-1)/2}
 H_{(d_s+n-1)/2}^{(1)}\!\left(\frac{\omega L_8^2}{r}\right),
 \qquad
 \operatorname{Re}\sigma(\omega)
 \propto\omega^{d_s+n-2}.
 \label{eq:IR_optical_Bessel}
\end{equation}
To obtain the power, expand the Hankel function at small
$\omega L_8^2/r$. Its two independent terms give an external
constant source and a normalizable response of order
$r^{-(d_s+n-1)}$. Their frequency ratio scales as
$\omega^{d_s+n-1}$.
The retarded current response acquires one fewer power in the
conductivity. Its spectral power equals that of the translationally invariant
parent Maxwell channel with the compact directions held
homogeneous. The derivation uses the local scaling region and
the effective five-dimensional gauge coupling. In a charged mixed
superfluid this gauge perturbation couples to scalar and metric
fluctuations.

\subsection{Analytic coupling thresholds and radial solvability}
\label{app:phase_criterion}
The quartic scalar coupling enters the eight-dimensional action
only through $(\boldsymbol\Phi^\dagger\boldsymbol\Phi)^2$. It consequently does not
change a neutral or normal-state solution, or either charged
linear onset equation. In the quartic functional it enters only
the scalar diagonal element. At a specified normal-state double
onset, the exact decomposition is
\begin{equation}
 b_\omega=b_\omega^{(0)},\qquad
 g=g^{(0)},\qquad
 b_\Phi=b_\Phi^{(0)}+\lambda_8 A_\Phi,
 \qquad A_\Phi>0.
 \label{eq:affine_quartic}
\end{equation}
The superscript zero means that the direct charged-scalar contact
term has been omitted. The other electric, supporting-field and
metric responses remain included. The coefficient $A_\Phi$ is the
positive norm of the quartic scalar profile with the inherited
radial measure. The determinant is therefore affine in $\lambda_8$
at the level of the action. For $b_\omega^{(0)}>0$,
\begin{align}
 \lambda_c&=
 \frac{(g^{(0)})^2-4b_\omega^{(0)}b_\Phi^{(0)}}{4b_\omega^{(0)}A_\Phi},
 \label{eq:symbolic_threshold}\\
 \det\mathbf B_\mu&=4A_\Phi\,b_\omega^{(0)}(\lambda_8-\lambda_c),
 \label{eq:affine_det}
\end{align}
which is Eq.~\eqref{eq:cartan_formal_threshold} at a general
double onset. The same construction applies to the canonical quartic
matrix after the rank-one correction \eqref{eq:canonical}, and
Eq.~\eqref{eq:threshold_shift} relates the two thresholds.

At the conformal double onset the profiles in
Eq.~\eqref{eq:zero_modes} allow the ultraviolet integrals to be
evaluated exactly. At the infrared endpoint the lowest
eigenfunctions are instead defined by Eqs.~\eqref{eq:IR_SL_p}
and~\eqref{eq:IR_SL_s}, the horizon regularity conditions
$w_1'(1)=0$ and $s_1'(1)=Ms_1(1)/7$, and the source-free
falloffs. Their overlap integrals
\eqref{eq:IR_direct_integrals} define an implicit infrared
threshold. Equation~\eqref{eq:threshold_shift} proves the
ordering of the fixed-density and fixed-potential thresholds
at each endpoint. It does not prove an ordering between the
ultraviolet and infrared thresholds. The possible interval
\eqref{eq:window} must therefore be checked within the same
parent theory before it can be used as a statement about
endpoint phase competition.

On a complete normal background with a common source-free
scalar mode $s_\star$, the direct charged-scalar contact term
has the exact coupling derivative
\begin{equation}
 \frac{\partial(\mathbf B_\mu)_{\Phi\Phi}}{\partial\lambda_8}
 =2A_\Phi^{\rm flow},\qquad
 A_\Phi^{\rm flow}=
 \frac14\int_{r_h}^\infty
 X(r)e^{4A(r)}s_\star(r)^4\,\dd r>0.
 \label{eq:full_contact_integral}
\end{equation}
The normal background and the charged linear zero modes do not
depend on $\lambda_8$, since that coupling first appears at
fourth order. Equation~\eqref{eq:full_contact_integral} therefore
establishes the affine dependence of the quartic determinant
on the parent coupling under the stated normal-state hypotheses.
It does not determine the complete-flow threshold and
does not establish that a full-flow double onset exists.

\subsubsection{Nonlinear charged profiles near the common onset}
The leading charged profiles are $w=w_0w_\star+\cdots$ and
$s=s_0s_\star+\cdots$. The neutral, electric and metric responses
begin at order $w_0^2$ and $s_0^2$, as in
Eq.~\eqref{eq:metric_response_expansion}. Solving their radial
equations with regular horizon data and fixed boundary sources
determines these induced profiles when no additional normalizable
zero mode is present.

Projection of the remaining field equations onto $w_\star$ and
$s_\star$ yields two amplitude equations. To leading order they are
\begin{equation}
 a_\omega+2b_\omega w_0^2+g\,s_0^2=0,
 \qquad
 a_\Phi+g\,w_0^2+2b_\Phi s_0^2=0.
 \label{eq:amplitude_equations}
\end{equation}
When the determinant in \eqref{eq:quartic} is positive, these
equations have a unique positive mixed solution in the wedge of
Figure~\ref{fig:phase}. The pure branches follow on its boundary.
The action evaluated on these branches gives the square differences
in Eq.~\eqref{eq:analytic_gaps}. The nonzero Jacobian in
\eqref{eq:critical} transfers this phase topology from $(a_\omega,a_\Phi)$
to the physical $(T,\mu)$ plane. If any of these hypotheses fails,
the shape of the local diagram must be determined by higher-order
terms or by a different onset structure.

\subsubsection{Direct evolution identities}
These identities give a direct implementation of the nonminimal
five-dimensional theory. Write $E=e^{-2A}$,
$U=-42/L_8^2+3m_\Psi^2f^2+3\lambda_\Psi f^4/2$ and
$\widetilde m^2=m_0^2+3\eta f^2$. Define
\begin{align}
 K_0&=\frac{3h}{X}f'^2+3k^2Xf^2,\nonumber\\
 K_s&=\frac hX s'^2+\frac E{4X}
       \left(-\frac{b^2}{h}+e^{-2B}w^2\right)s^2,\nonumber\\
 G^2&=\frac2{X^2}
 \left[-Eb'^2+hEe^{-2B}w'^2-\frac{E^2e^{-2B}}h b^2w^2\right].
\end{align}
Combining the trace of \eqref{eq:8dEinstein} with the charged scalar equation eliminates second derivatives from the eight-dimensional Ricci scalar,
\begin{equation}
 R_8=
 \frac{3K_0+4U+\eps\left[
 G^2/2+(3/2-7\xi)K_s+(2-7\xi)\widetilde m^2s^2
 +(1-7\xi)\lambda_8s^4\right]}
 {3+\eps\xi(7\xi-3/2)s^2}.
 \label{eq:Rtrace}
\end{equation}
The matter equations are
\begin{align}
 f''&=-\left(4A'+B'+\frac{h'}h\right)f'
       +\frac{V_f}{6h}+\frac{\eps\eta X}{2h}fs^2,\nonumber\\
 b''&=-\left(2A'+B'-\frac{\phi'}{\sqrt3}\right)b'
       +\left(\frac{Ee^{-2B}}h w^2+\frac X{4h}s^2\right)b,\nonumber\\
 w''&=-\left(2A'-B'+\frac{h'}h-\frac{\phi'}{\sqrt3}\right)w'
       -\frac E{h^2}b^2w+\frac X{4h}s^2w,\nonumber\\
 s''&=-\left(4A'+B'+\frac{h'}h\right)s'
       +\left[\frac Xh(\widetilde m^2+\xi R_8+\lambda_8s^2)
       -\frac E{4h^2}b^2+\frac{Ee^{-2B}}{4h}w^2\right]s.
 \label{eq:matterODE}
\end{align}
In particular, the curvature contribution is computed from the charged geometry, not inserted from a neutral probe background.

The metric function $B(r)$ obeys the radial equation and horizon condition
\begin{align}
 B''={}&-\left(4A'+B'+\frac{h'}h+\frac{\mathcal F'}{\mathcal F}\right)B'
 -\frac{\eps}{2\mathcal F}\left[
 \frac{Ee^{-2B}}X w'^2-\frac{E^2e^{-2B}}{Xh^2}b^2w^2
 +\frac{Ee^{-2B}}{4h}w^2s^2\right],\\
 B'_h={}&-\frac{\eps e^{-2A_h-2B_h}w_h^2s_h^2}{8\mathcal F_h h'_h}.
\end{align}
The regular horizon conditions also include
\begin{equation}
 w'_h=\frac{X_hs_h^2}{4h'_h}w_h,\qquad
 s'_h=\frac{X_h(\widetilde m_h^2+\xi R_{8,h}+\lambda_8s_h^2)
            +e^{-2A_h-2B_h}w_h^2/4}{h'_h}s_h.
\end{equation}
The scalar quartic term is retained here. These equations and the independent tensor tests make the nonminimal backreaction explicit.


\begin{thebibliography}{99}\raggedright\setlength{\itemsep}{2pt}

\bibitem{Maldacena:1997re}
J.~M.~Maldacena,
``The Large $N$ limit of superconformal field theories and supergravity,''
Adv. Theor. Math. Phys. \textbf{2} (1998) 231--252,
\url{https://doi.org/10.4310/ATMP.1998.v2.n2.a1}
[hep-th/9711200].

\bibitem{Gubser:1998bc}
S.~S.~Gubser, I.~R.~Klebanov and A.~M.~Polyakov,
``Gauge theory correlators from noncritical string theory,''
Phys. Lett. B \textbf{428} (1998) 105--114,
\url{https://doi.org/10.1016/S0370-2693(98)00377-3}
[hep-th/9802109].

\bibitem{Witten:1998qj}
E.~Witten,
``Anti de Sitter space and holography,''
Adv. Theor. Math. Phys. \textbf{2} (1998) 253--291,
\url{https://doi.org/10.4310/ATMP.1998.v2.n2.a2}
[hep-th/9802150].

\bibitem{Aharony:1999ti}
O.~Aharony, S.~S.~Gubser, J.~M.~Maldacena, H.~Ooguri and Y.~Oz,
``Large N field theories, string theory and gravity,''
Phys. Rept. \textbf{323} (2000) 183--386,
\url{https://doi.org/10.1016/S0370-1573(99)00083-6}
[hep-th/9905111].

\bibitem{Hartnoll:2009sz}
S.~A.~Hartnoll,
``Lectures on holographic methods for condensed matter physics,''
Class. Quant. Grav. \textbf{26} (2009) 224002,
\url{https://doi.org/10.1088/0264-9381/26/22/224002}
[arXiv:0903.3246 [hep-th]].

\bibitem{McGreevy:2009xe}
J.~McGreevy,
``Holographic duality with a view toward many-body physics,''
Adv. High Energy Phys. \textbf{2010} (2010) 723105,
\url{https://doi.org/10.1155/2010/723105}
[arXiv:0909.0518 [hep-th]].

\bibitem{Herzog:2009xv}
C.~P.~Herzog,
``Lectures on Holographic Superfluidity and Superconductivity,''
J. Phys. A \textbf{42} (2009) 343001,
\url{https://doi.org/10.1088/1751-8113/42/34/343001}
[arXiv:0904.1975 [hep-th]].

\bibitem{Horowitz:2010gk}
G.~T.~Horowitz,
``Introduction to Holographic Superconductors,''
Lect. Notes Phys. \textbf{828} (2011) 313--347,
\url{https://doi.org/10.1007/978-3-642-04864-7_10}
[arXiv:1002.1722 [hep-th]].

\bibitem{Cai:2015cya}
R.-G.~Cai, L.~Li, L.-F.~Li and R.-Q.~Yang,
``Introduction to Holographic Superconductor Models,''
Sci. China Phys. Mech. Astron. \textbf{58} (2015) 060401,
\url{https://doi.org/10.1007/s11433-015-5676-5}
[arXiv:1502.00437 [hep-th]].

\bibitem{Gubser:2008px}
S.~S.~Gubser,
``Breaking an Abelian gauge symmetry near a black hole horizon,''
Phys. Rev. D \textbf{78} (2008) 065034,
\url{https://doi.org/10.1103/PhysRevD.78.065034}
[arXiv:0801.2977 [hep-th]].

\bibitem{Hartnoll:2008vx}
S.~A.~Hartnoll, C.~P.~Herzog and G.~T.~Horowitz,
``Building a Holographic Superconductor,''
Phys. Rev. Lett. \textbf{101} (2008) 031601,
\url{https://doi.org/10.1103/PhysRevLett.101.031601}
[arXiv:0803.3295 [hep-th]].

\bibitem{Hartnoll:2008kx}
S.~A.~Hartnoll, C.~P.~Herzog and G.~T.~Horowitz,
``Holographic Superconductors,''
JHEP \textbf{12} (2008) 015,
\url{https://doi.org/10.1088/1126-6708/2008/12/015}
[arXiv:0810.1563 [hep-th]].

\bibitem{Franco:2009if}
S.~Franco, A.~M.~Garcia-Garcia and D.~Rodriguez-Gomez,
``A General class of holographic superconductors,''
JHEP \textbf{04} (2010) 092,
\url{https://doi.org/10.1007/JHEP04(2010)092}
[arXiv:0906.1214 [hep-th]].

\bibitem{Gubser:2008wv}
S.~S.~Gubser,
``Colorful horizons with charge in anti-de Sitter space,''
Phys. Rev. Lett. \textbf{101} (2008) 191601,
\url{https://doi.org/10.1103/PhysRevLett.101.191601}
[arXiv:0803.3483 [hep-th]].

\bibitem{Gubser:2008zu}
S.~S.~Gubser and S.~S.~Pufu,
``The Gravity dual of a p-wave superconductor,''
JHEP \textbf{11} (2008) 033,
\url{https://doi.org/10.1088/1126-6708/2008/11/033}
[arXiv:0805.2960 [hep-th]].

\bibitem{Ammon:2010pg}
M.~Ammon, J.~Erdmenger, V.~Grass, P.~Kerner and A.~O'Bannon,
``On Holographic p-wave Superfluids with Back-reaction,''
Phys. Lett. B \textbf{686} (2010) 192--198,
\url{https://doi.org/10.1016/j.physletb.2010.02.021}
[arXiv:0912.3515 [hep-th]].

\bibitem{HerzogPufu}
C.~P.~Herzog and S.~S.~Pufu,
``The Second Sound of SU(2),''
JHEP \textbf{04} (2009) 126,
\url{https://doi.org/10.1088/1126-6708/2009/04/126}
[arXiv:0902.0409 [hep-th]].

\bibitem{Basu:2011tt}
P.~Basu and J.~H.~Oh,
``Analytic Approaches to Anisotropic Holographic Superfluids,''
JHEP \textbf{07} (2012) 106,
\url{https://doi.org/10.1007/JHEP07(2012)106}
[arXiv:1109.4592 [hep-th]].

\bibitem{Herzog2010}
C.~P.~Herzog,
``An Analytic Holographic Superconductor,''
Phys. Rev. D \textbf{81} (2010) 126009,
\url{https://doi.org/10.1103/PhysRevD.81.126009}
[arXiv:1003.3278 [hep-th]].

\bibitem{Basu2010}
P.~Basu, J.~He, A.~Mukherjee, M.~Rozali and H.-H.~Shieh,
``Competing Holographic Orders,''
JHEP \textbf{10} (2010) 092,
\url{https://doi.org/10.1007/JHEP10(2010)092}
[arXiv:1007.3480 [hep-th]].

\bibitem{Amado:2013splusp}
I.~Amado, D.~Arean, A.~Jimenez-Alba, L.~Melgar and I.~Salazar Landea,
``Holographic s+p Superconductors,''
Phys. Rev. D \textbf{89} (2014) 026009,
\url{https://doi.org/10.1103/PhysRevD.89.026009}
[arXiv:1309.5086 [hep-th]].

\bibitem{Nie:2013sda}
Z.-Y.~Nie, R.-G.~Cai, X.~Gao and H.~Zeng,
``Competition between the s-wave and p-wave superconductivity phases in a holographic model,''
JHEP \textbf{11} (2013) 087,
\url{https://doi.org/10.1007/JHEP11(2013)087}
[arXiv:1309.2204 [hep-th]].

\bibitem{Nie:2014backreaction}
Z.-Y.~Nie, R.-G.~Cai, X.~Gao, L.~Li and H.~Zeng,
``Phase transitions in a holographic s  $+$  p model with back-reaction,''
Eur. Phys. J. C \textbf{75} (2015) 559,
\url{https://doi.org/10.1140/epjc/s10052-015-3773-2}
[arXiv:1501.00004 [hep-th]].

\bibitem{Nishida:2014lta}
M.~Nishida,
``Phase Diagram of a Holographic Superconductor Model with s-wave and d-wave,''
JHEP \textbf{09} (2014) 154,
\url{https://doi.org/10.1007/JHEP09(2014)154}
[arXiv:1403.6070 [hep-th]].

\bibitem{Li:2014wca}
L.-F.~Li, R.-G.~Cai, L.~Li and Y.-Q.~Wang,
``Competition between s-wave order and d-wave order in holographic superconductors,''
JHEP \textbf{08} (2014) 164,
\url{https://doi.org/10.1007/JHEP08(2014)164}
[arXiv:1405.0382 [hep-th]].

\bibitem{Yin:2013}
L.~Yin, D.~Hou and H.-c.~Ren,
``Ginzburg-Landau theory of a holographic superconductor,''
Phys. Rev. D \textbf{91} (2015) 026003,
\url{https://doi.org/10.1103/PhysRevD.91.026003}
[arXiv:1311.3847 [hep-th]].

\bibitem{Natsuume:2025GL}
M.~Natsuume,
``What is the Dual Ginzburg--Landau Theory for Holographic Superconductors?,''
Prog. Theor. Exp. Phys. \textbf{2025} (2025) 023B08,
\url{https://doi.org/10.1093/ptep/ptaf018}
[arXiv:2407.13956 [hep-th]].

\bibitem{Zhao:2023stability}
Z.-Q.~Zhao, X.-K.~Zhang and Z.-Y.~Nie,
``Dynamical stability from quasi normal modes in 2nd, 1st and 0th order
holographic superfluid phase transitions,''
JHEP \textbf{02} (2023) 023,
\url{https://doi.org/10.1007/JHEP02(2023)023}.

\bibitem{Cao:2024susceptibility}
Y.-X.~Cao, H.~Zeng and Z.-Y.~Nie,
``The local thermodynamic instability from negative susceptibility in a
holographic superfluid with nonlinear terms,''
[arXiv:2409.14407 [hep-th]].

\bibitem{Zhao:2024spinodal}
X.~Zhao, Z.-Y.~Nie, Z.-Q.~Zhao, H.-B.~Zeng, Y.~Tian and M.~Baggioli,
``Dynamical evolution of spinodal decomposition in holographic superfluids,''
JHEP \textbf{02} (2024) 184,
\url{https://doi.org/10.1007/JHEP02(2024)184}.

\bibitem{BreitenlohnerFreedman:1982}
P.~Breitenlohner and D.~Z.~Freedman,
``Stability in Gauged Extended Supergravity,''
Ann. Phys. \textbf{144} (1982) 249,
\url{https://doi.org/10.1016/0003-4916(82)90116-6}
.

\bibitem{KlebanovWitten:1999}
I.~R.~Klebanov and E.~Witten,
``AdS / CFT correspondence and symmetry breaking,''
Nucl. Phys. B \textbf{556} (1999) 89--114,
\url{https://doi.org/10.1016/S0550-3213(99)00387-9}
[hep-th/9905104].

\bibitem{Bianchi:2001kw}
M.~Bianchi, D.~Z.~Freedman and K.~Skenderis,
``Holographic renormalization,''
Nucl. Phys. B \textbf{631} (2002) 159--194,
\url{https://doi.org/10.1016/S0550-3213(02)00179-7}
[hep-th/0112119].

\bibitem{Skenderis:2002wp}
K.~Skenderis,
``Lecture notes on holographic renormalization,''
Class. Quant. Grav. \textbf{19} (2002) 5849--5876,
\url{https://doi.org/10.1088/0264-9381/19/22/306}
[hep-th/0209067].

\bibitem{Kachru:2008yh}
S.~Kachru, X.~Liu and M.~Mulligan,
``Gravity duals of Lifshitz-like fixed points,''
Phys. Rev. D \textbf{78} (2008) 106005,
\url{https://doi.org/10.1103/PhysRevD.78.106005}
[arXiv:0808.1725 [hep-th]].

\bibitem{Charmousis:2010zz}
C.~Charmousis, B.~Gouteraux, B.~S.~Kim, E.~Kiritsis and R.~Meyer,
``Effective Holographic Theories for low-temperature condensed matter systems,''
JHEP \textbf{11} (2010) 151,
\url{https://doi.org/10.1007/JHEP11(2010)151}
[arXiv:1005.4690 [hep-th]].

\bibitem{Gouteraux:2012yr}
B.~Gout\'{e}raux and E.~Kiritsis,
``Generalized Holographic Quantum Criticality at Finite Density,''
JHEP \textbf{12} (2011) 036,
\url{https://doi.org/10.1007/JHEP12(2011)036}
[arXiv:1107.2116 [hep-th]].

\bibitem{Gouteraux:2011ce}
B.~Gouteraux and E.~Kiritsis,
``Quantum critical lines in holographic phases with (un)broken symmetry,''
JHEP \textbf{04} (2013) 053,
\url{https://doi.org/10.1007/JHEP04(2013)053}
[arXiv:1212.2625 [hep-th]].

\bibitem{Huijse:2011ef}
L.~Huijse, S.~Sachdev and B.~Swingle,
``Hidden Fermi surfaces in compressible states of gauge-gravity duality,''
Phys.Rev.B \textbf{85} (2012) 035121,
\url{https://doi.org/10.1103/PhysRevB.85.035121}
[arXiv:1112.0573 [cond-mat.str-el]].

\bibitem{Dong:2012se}
X.~Dong, S.~Harrison, S.~Kachru, G.~Torroba and H.~Wang,
``Aspects of holography for theories with hyperscaling violation,''
JHEP \textbf{06} (2012) 041,
\url{https://doi.org/10.1007/JHEP06(2012)041}
[arXiv:1201.1905 [hep-th]].

\bibitem{Alishahiha:2012qu}
M.~Alishahiha, E.~O.~Colgain and H.~Yavartanoo,
``Charged Black Branes with Hyperscaling Violating Factor,''
JHEP \textbf{11} (2012) 137,
\url{https://doi.org/10.1007/JHEP11(2012)137}
[arXiv:1209.3946 [hep-th]].

\bibitem{GubserNellore:2008}
S.~S.~Gubser and A.~Nellore,
``Mimicking the QCD equation of state with a dual black hole,''
Phys. Rev. D \textbf{78} (2008) 086007,
\url{https://doi.org/10.1103/PhysRevD.78.086007}
[arXiv:0804.0434 [hep-th]].

\bibitem{Arefeva:2018}
I.~Ya.~Aref'eva, A.~A.~Golubtsova and G.~Policastro,
``Exact holographic RG flows and the $A_1\times A_1$ Toda chain,''
JHEP \textbf{05} (2019) 117,
\url{https://arxiv.org/abs/1803.06764}.

\bibitem{CremoniniLi:2016}
S.~Cremonini and L.~Li,
``Criteria for superfluid instabilities of geometries with hyperscaling violation,''
JHEP \textbf{11} (2016) 137,
\url{https://arxiv.org/abs/1606.02745}.

\bibitem{RenXie:2023}
J.~Ren and H.~Xie,
``Holographic superconductors at zero density,''
Phys. Rev. D \textbf{107} (2023) 106004,
\url{https://arxiv.org/abs/2206.03498}.

\bibitem{Giataganas:2025}
D.~Giataganas, U.~G\"ursoy, C.~Moran, J.~F.~Pedraza
and D.~Rodr\'iguez Fern\'andez,
``Anisotropic critical points from holography,''
JHEP \textbf{03} (2026) 026,
\url{https://arxiv.org/abs/2509.15021}.

\bibitem{Park:2016gzx}
M.~Park, J.~Park and J.~H.~Oh,
``Phase transition in anisotropic holographic superfluids with arbitrary dynamical critical exponent $z$ and hyperscaling violation factor $\alpha $,''
Eur. Phys. J. C \textbf{77} (2017) 810,
\url{https://doi.org/10.1140/epjc/s10052-017-5382-8}
[arXiv:1609.08241 [hep-th]].

\bibitem{Pan:2015hsv}
Q.~Pan and S.-J.~Zhang,
``Revisiting holographic superconductors with hyperscaling violation,''
Eur. Phys. J. C \textbf{76} (2016) 126,
\url{https://doi.org/10.1140/epjc/s10052-016-3980-5}
[arXiv:1510.09199 [hep-th]].

\bibitem{Kim:2025hsv}
G.~Kim, Y.~S.~Choi and J.~H.~Oh,
``Analytic approaches to anisotropic holographic superfluids in asymptotically hyperscaling violation geometry,''
[arXiv:2504.13635 [hep-th]].

\bibitem{Ryu:2006bv}
S.~Ryu and T.~Takayanagi,
``Holographic derivation of entanglement entropy from AdS/CFT,''
Phys. Rev. Lett. \textbf{96} (2006) 181602,
\url{https://doi.org/10.1103/PhysRevLett.96.181602}
[hep-th/0603001].

\bibitem{Ryu:2006ef}
S.~Ryu and T.~Takayanagi,
``Aspects of Holographic Entanglement Entropy,''
JHEP \textbf{08} (2006) 045,
\url{https://doi.org/10.1088/1126-6708/2006/08/045}
[hep-th/0605073].

\bibitem{Casini:2011kv}
H.~Casini, M.~Huerta and R.~C.~Myers,
``Towards a derivation of holographic entanglement entropy,''
JHEP \textbf{05} (2011) 036,
\url{https://doi.org/10.1007/JHEP05(2011)036}
[arXiv:1102.0440 [hep-th]].

\bibitem{Bhattacharya:2012mi}
J.~Bhattacharya, M.~Nozaki, T.~Takayanagi and T.~Ugajin,
``Thermodynamical Property of Entanglement Entropy for Excited States,''
Phys. Rev. Lett. \textbf{110} (2013) 091602,
\url{https://doi.org/10.1103/PhysRevLett.110.091602}
[arXiv:1212.1164 [hep-th]].

\bibitem{Cai:2012nm}
R.-G.~Cai, S.~He, L.~Li and Y.-L.~Zhang,
``Holographic Entanglement Entropy on P-wave Superconductor Phase Transition,''
JHEP \textbf{07} (2012) 027,
\url{https://doi.org/10.1007/JHEP07(2012)027}
[arXiv:1204.5962 [hep-th]].

\bibitem{Park:2022oek}
C.~Park, G.~Kim, J.-s.~Chae and J.~H.~Oh,
``Holographic entanglement entropy probe on spontaneous symmetry breaking with vector order,''
JHEP \textbf{02} (2023) 182,
\url{https://doi.org/10.1007/JHEP02(2023)182}
[arXiv:2210.08919 [hep-th]].

\bibitem{Chen:2024isotropy}
C.-Y.~Chen, M.-J.~Li, Z.~Yang, D.-M.~Jin and P.~Liu,
``Diagnosing emergent isotropy in anisotropic holographic systems using quantum information measures,''
Eur. Phys. J. C \textbf{85} (2025) 136,
\url{https://doi.org/10.1140/epjc/s10052-025-13797-8}
[arXiv:2404.11021 [hep-th]].

\bibitem{Martini:2018}
R.~Martini and O.~Zanusso,
``Renormalization of multicritical scalar models in curved space,''
\url{https://doi.org/10.1140/epjc/s10052-019-6721-8}
[arXiv:1810.06395 [hep-th]].



\bibitem{GouterauxSmolic:2012}
B.~Gout\'eraux, J.~Smolic, M.~Smolic, K.~Skenderis and M.~Taylor,
``Holography for Einstein-Maxwell-dilaton theories from generalized
dimensional reduction,'' JHEP \textbf{01} (2012) 089,
\url{https://arxiv.org/abs/1110.2320}.

\bibitem{Smolic:2013}
M.~Smolic, ``Holography and hydrodynamics for EMD theory with two Maxwell
fields,'' JHEP \textbf{03} (2013) 124,
\url{https://arxiv.org/abs/1301.6020}.

\bibitem{Charmousis:2012reduction}
C.~Charmousis, B.~Gout\'eraux and E.~Kiritsis,
``Higher-derivative scalar-vector-tensor theories: black holes, Galileons,
singularity cloaking and holography,'' JHEP \textbf{09} (2012) 011,
\url{https://arxiv.org/abs/1206.1499}.

\bibitem{Gath:2013}
J.~Gath, J.~Hartong, R.~Monteiro and N.~A.~Obers,
``Holographic models for theories with hyperscaling violation,''
JHEP \textbf{04} (2013) 159,
\url{https://arxiv.org/abs/1212.3263}.

\bibitem{Perlmutter:2012}
E.~Perlmutter, ``Hyperscaling violation from supergravity,''
JHEP \textbf{06} (2012) 165,
\url{https://arxiv.org/abs/1205.0242}.

\bibitem{CremoniniSinkovics:2014}
S.~Cremonini and A.~Sinkovics,
``Spatially modulated instabilities of geometries with hyperscaling
violation,'' JHEP \textbf{01} (2014) 099,
\url{https://arxiv.org/abs/1212.4172}.

\bibitem{DonosGauntlett:2014}
A.~Donos and J.~P.~Gauntlett, ``Holographic Q-lattices,''
JHEP \textbf{04} (2014) 040,
\url{https://arxiv.org/abs/1311.3292}.

\bibitem{CordovaDeLucaTomasiello:2019}
C.~C\'ordova, G.~B.~De Luca and A.~Tomasiello,
``AdS$_8$ solutions in type II supergravity,''
JHEP \textbf{07} (2019) 127,
\url{https://doi.org/10.1007/JHEP07(2019)127}
[arXiv:1811.06987 [hep-th]].

\bibitem{AnastasiouAraya:2026}
G.~Anastasiou, I.~J.~Araya, N.~Boulanger, R.~Olea and D.~Rovere,
``Conformal Renormalisation of 8D Einstein Gravity,''
[arXiv:2607.03679 [hep-th]],
\url{https://arxiv.org/abs/2607.03679}.

\bibitem{GKN:2024}
A.~Ghodsi, E.~Kiritsis and F.~Nitti,
``On holographic confining QFTs on AdS,''
[arXiv:2409.02879 [hep-th]],
\url{https://arxiv.org/abs/2409.02879}.

\bibitem{KKN:2025}
J.~Kastikainen, E.~Kiritsis and F.~Nitti,
``Holographic confining theories on space-times with constant positive curvature,''
[arXiv:2502.04036 [hep-th]],
\url{https://arxiv.org/abs/2502.04036}.

\bibitem{GKN:2026}
A.~Ghodsi, E.~Kiritsis and F.~Nitti,
``The phase diagram of confining holographic theories on constant
curvature manifolds in the presence of a $\theta$-angle,''
[arXiv:2604.20418 [hep-th]],
\url{https://arxiv.org/abs/2604.20418}.

\bibitem{JainJarvinen:2025}
P.~Jain and M.~J\"arvinen,
``An analytic approach for holographic entanglement entropy at
(quantum) criticality,''
[arXiv:2509.25360 [hep-th]],
\url{https://arxiv.org/abs/2509.25360}.

\bibitem{Faedo:2024}
A.~F.~Faedo, C.~Hoyos, M.~Piai, J.~Rodgers and J.~G.~Subils,
``Light holographic dilatons near critical points,''
[arXiv:2406.04974 [hep-th]],
\url{https://arxiv.org/abs/2406.04974}.

\bibitem{Bobev:2025Dp}
N.~Bobev, G.~Mera \'Alvarez and H.~Paul,
``Correlation functions for non-conformal D$p$-brane holography,''
JHEP \textbf{07} (2025) 137,
\url{https://arxiv.org/abs/2503.18770}.

\bibitem{Bobev:2026Scale}
N.~Bobev, P.~Bomans and F.~F.~Gautason,
``Scale-Covariant Holography,''
[arXiv:2609.38302 [hep-th]],
\url{https://arxiv.org/abs/2609.38302}.

\bibitem{Conti:2026D2}
A.~Conti, A.~Guarino and R.~Stuardo,
``Scaling similarity and generalised conformal symmetry in D2-brane holography,''
[arXiv:2609.38336 [hep-th]],
\url{https://arxiv.org/abs/2609.38336}.


\end{thebibliography}
\end{document}